\documentclass[prd,showpacs,showkeys,nofootinbib,twocolumn, superscriptaddress]{revtex4-2}
\usepackage[normalem]{ulem}
\usepackage{amsmath}
\usepackage{graphicx}
\usepackage[colorlinks=true, allcolors=blue, citecolor=blue,      
    urlcolor=blue]{hyperref}

\begin{document}

\title{Cosmological evolution of interacting dark energy with a CPL equation of state}
\author{Gerald Neumann}
\email[Contact author: ]{gerald.neumann.d@mail.pucv.cl}
\email{gneuman@usm.cl}
\affiliation{Instituto de F\'isica, Pontificia Universidad Cat\'olica de Valpara\'iso, Avenida Universidad
330, Valpara\'iso, Chile}
\affiliation{Departamento de F\'isica, Universidad T\'ecnica Federico Santa Mar\'ia,  Av. España 1680, Casilla 110-V, Valpara\'iso, Chile.}

\author{Dorian Araya}
\email[Contact author: ]{dorian.araya.a@mail.pucv.cl}
\affiliation{Instituto de F\'isica, Pontificia Universidad Cat\'olica de Valpara\'iso, Avenida Universidad
330, Valpara\'iso, Chile}

\author{Nelson Videla}
\email[Contact author: ]{nelson.videla@pucv.cl}
\affiliation{Instituto de F\'isica, Pontificia Universidad Cat\'olica de Valpara\'iso, Avenida Universidad
330, Valpara\'iso, Chile}

\date{ \today}

\begin{abstract}
This paper examines interacting dark energy models within the Chevallier–Polarski–Linder (CPL) parametrization, emphasizing both theoretical structure and observational viability. Two commonly adopted interaction terms are considered: $Q = \beta H \rho_\text{de}$ and $Q = \beta H \rho_\text{c}$. We derive exact analytic solutions that describe how the dark sector evolves. These solutions involve incomplete gamma functions and reveal a non-trivial mathematical structure that is often missed in numerical analyses. We perform a Bayesian analysis using current cosmological observations, including the Hubble parameter (OHD), Type Ia supernovae (SNIa), baryon acoustic oscillations (BAO), and cosmic microwave background (CMB) data. Relative to the non-interacting CPL scenario, the interacting model with $Q = \beta H \rho_\text{de}$ yields a modestly improved fit, as indicated by the Akaike Information Criterion (AIC). However, the Bayesian Information Criterion (BIC) penalizes increased model complexity, leading to a continued preference for $\Lambda$CDM. In contrast, the interaction model that depends on dark matter density does not provide observational support. The preferred interacting scenario indicates that the dark energy equation of state evolves dynamically, transitioning from an effective phantom regime at high redshift to quintessence-like behavior at late times. Further analysis suggests the possibility of a future transition from accelerated to decelerated expansion, primarily associated with the reconstructed CPL dynamics, while the interaction alters the detailed evolution of the dark sector and the precise transition epoch. These findings suggest that interacting dark energy models within the CPL framework enrich the standard cosmological model by introducing more diverse phenomenology while maintaining consistency with current observations.
\end{abstract}
\maketitle

\section{Introduction}
The accelerated expansion of the late-time Universe is supported by several independent cosmological observations, including Type Ia supernovae (SNIa) \cite{SupernovaSearchTeam:1998fmf, SupernovaCosmologyProject:1998vns, Pan-STARRS1:2017jku}, cosmic microwave background anisotropies (CMB) \cite{Planck:2018vyg}, and baryon acoustic oscillations (BAO) in large-scale structure surveys \cite{SDSS:2005xqv, eBOSS:2020yzd, DESI:2024mwx}. According to General Relativity (GR), this is explained by introducing a negative-pressure component in the energy–momentum tensor, commonly known as dark energy (DE) \cite{Copeland:2006wr}. In the standard $\Lambda$CDM model, dark energy is identified with a cosmological constant $\Lambda$, contributing approximately $70\%$ of the total energy density of the Universe, while cold dark matter (CDM) accounts for about $25\%$, and baryonic matter for the remaining $5\%$ \cite{Planck:2018vyg}. Despite its remarkable success in describing a wide range of cosmological observations, the $\Lambda$CDM paradigm faces several theoretical and observational challenges.

The cosmological constant problem, from a theoretical standpoint, stems from the substantial discrepancy between the observed value of $\Lambda$ and the vacuum energy density predicted by quantum field theory \cite{Weinberg:1988cp,Martin:2012bt}. Observationally, notable tensions have emerged in recent years. Specifically, the Hubble constant $H_0$ inferred from early-Universe measurements by the \textit{Planck} Collaboration \cite{Planck:2018vyg} exhibits a $4$ to $6\sigma$ tension with late-time determinations based on Type Ia supernovae \cite{Riess:2019cxk, Riess:2021jrx}. Furthermore, a discrepancy in the amplitude of matter fluctuations, quantified by the parameter $S_8 \equiv \left(\frac{\Omega_{\mathrm{m}}}{0.3}\right)^{1/2}\sigma_8$, has been identified, with low-redshift probes consistently yielding lower values than those inferred from cosmic microwave background (CMB) observations \cite{DES:2017myr,Heymans:2020gsg}. For recent reviews on cosmological tensions and their implications, see Refs.~\cite{Verde:2019ivm,DiValentino:2020zio,DiValentino:2020vvd,DiValentino:2021izs,Perivolaropoulos:2021jda}

These tensions, together with the conceptual challenges associated with the cosmological constant, have motivated the exploration of extensions to the $\Lambda$CDM model. Broadly, such extensions can be classified into two categories: modifications of GR, and models involving dynamical dark energy, in which the properties of dark energy evolve with time, in contrast to the constant $\Lambda$ scenario. A wide variety of dynamical dark energy candidates have been proposed and extensively studied in the literature. For comprehensive reviews, see Refs.~\cite{Copeland:2006wr,Nojiri:2006ri,Clifton:2011jh,Bamba2012,Joyce:2016vqv}. A common approach to describing the possible time evolution of dark energy is through phenomenological parametrizations of its equation of state. Among these, the Chevallier–Polarski–Linder (CPL) parametrization \cite{Chevallier:2000qy,Linder:2002et},
\begin{equation}
    w_\text{de}(z) = w_0 + w_a \frac{z}{1+z},\label{CPL eos}
\end{equation}
provides a simple yet flexible framework for describing deviations from a constant equation of state, while remaining well behaved over the relevant redshift range. Here, $w_0$ represents the value reached at present times, while $w_a$ modulates the time evolution in the past.

Recent observational results, particularly from the Dark Energy Spectroscopic Instrument (DESI) \cite{DESI:2024mwx, DESI:2025zgx}, indicate increased sensitivity to mild deviations from $w_\text{de}=-1$ and provide possible evidence for evolving dark energy \cite{DESI:2024aqx,Malekjani:2024bgi,Du:2024pai,Giare:2024oil,Zheng:2024qzi,Ormondroyd:2025iaf,Nesseris:2025lke,DESI:2025wyn, Scherer:2025esj,Capozziello:2025qmh}. In this context, parametrizations such as CPL play a central role in interpreting these deviations and quantifying potential departures from a cosmological constant. In addition, recent analyses suggest that the observed preference for phantom dark energy in CPL fits to current data may be an effective phenomenon resulting from interactions within the dark sector, rather than a fundamental violation of standard energy conditions \cite{Guedezounme:2025wav}.

Moreover, there is no fundamental reason to assume that dark matter and dark energy evolve independently. This has motivated the study of interacting dark sector models, in which an energy exchange between dark matter and dark energy is allowed, providing a natural extension of the standard cosmological framework \cite{Wetterich:1994bg,Amendola:1999er,Zimdahl:2001ar,Chimento:2003iea,Farrar:2003uw}. Such interactions can modify both the background expansion history and the growth of cosmic structures, leading to observable signatures at different cosmological scales \cite{Valiviita:2008iv,He:2008si,HeWang2008,Jackson:2009mz,Bolotin2015,Wang2016}. Consequently, interacting dark energy models have been extensively investigated as a possible mechanism to alleviate current cosmological tensions and to provide a more general description of the dark sector \cite{DiValentino:2017iww,Yang:2018uae,Yang:2018euj,Pan:2019gop,DiValentino:2019ffd,Wang:2024vmw}.

A wide range of phenomenological interaction models have been proposed in the literature. Among the most commonly studied cases are interaction terms proportional to the dark matter density, $Q \propto H\rho_\text{c}$, and to the dark energy density, $Q \propto H\rho_\text{de}$ \cite{Amendola:1999er,Zimdahl:2001ar,Chimento:2003iea,Valiviita:2008iv,vanderWesthuizen:2025vcb}. More general forms, including linear combinations such as $Q \propto H(\alpha \rho_\text{c} + \beta \rho_\text{de})$ and nonlinear interactions, have also been extensively explored \cite{Gavela:2009cy,Costa:2013sva,Pan:2012ki,vanderWesthuizen:2025mnw,vanderWesthuizen:2025rip}.
Foundational studies demonstrated that these models may exhibit large-scale instabilities at the perturbation level unless both the interaction and momentum transfer are consistently defined \cite{Valiviita:2008iv,He:2008si}. Consequently, perturbative stability has become a central theoretical criterion for the viability of interacting scenarios. Subsequent developments have introduced more advanced perturbative frameworks and have systematically examined observational constraints, parameter degeneracies, and consistency conditions for interacting models \cite{Yang:2018euj,Clemson:2011an,Li:2014eha}. In recent years, interacting dark energy scenarios have been re-examined in the context of high-precision cosmological data, such as DESI measurements, which has led to renewed interest in their phenomenology and viability \cite{vanderWesthuizen:2023hcl,Hoerning:2023hks,Giare:2024smz,Benisty:2024lmj,Ghedini:2024mdu,Zhu:2025lrk,Tsedrik:2025jdv,Petri:2025swg,Figueruelo:2026eis,Li:2026xaz}. These investigations demonstrate that interacting models can yield fits to current datasets that are competitive with, or superior to, those of $\Lambda$CDM or dynamical dark energy parametrizations, while also generating novel physical effects at both the background and perturbation levels.

Further studies have shown that dark-sector interactions can be partially degenerate with a time-dependent dark energy equation of state at the background level, see e.g. \cite{Clemson:2011an,Xu:2011tsa,Carneiro:2014uua}. This degeneracy persists when specific parametrizations, such as CPL, are adopted, since different combinations of interaction and $w_\text{de}(z)$ can lead to nearly identical expansion histories, making them difficult to distinguish using background observables alone. In the DESI era, this degeneracy has become particularly relevant, as interacting models can mimic evolving dark energy at the background level. It has recently been shown that this degeneracy can be made exact for CPL-like parametrizations through a mapping between the interaction term and an effective equation of state \cite{Petri:2025swg}, and can only be broken at the perturbation level via its impact on structure formation and CMB observables. In this context, interacting CPL models can therefore be classified into degenerate constructions and genuinely dynamical scenarios, the latter being the focus of this work.

Unlike previous approaches, this work does not impose a background-level degeneracy between interacting and non-interacting models \cite{Clemson:2011an,Xu:2011tsa,Carneiro:2014uua}.

The novelty of this study lies not only in confronting interacting CPL models with observational data, but also in deriving exact analytical solutions for dark-sector evolution in two widely studied interacting scenarios. These solutions reveal a non-trivial mathematical structure involving incomplete gamma functions and provide a direct analytical understanding of the impact of dark-sector interactions on cosmological dynamics. Consequently, the present analysis complements the predominantly numerical literature, see e.g.\cite{Yang:2018uae,Pan:2019gop}, and establishes a transparent connection between the mathematical properties of the solutions and their observational implications.

This study examines phenomenological interaction terms and directly evaluates their influence on the expansion history and dark-energy dynamics within the CPL framework, thereby enabling genuinely distinct cosmological evolutions. In contrast to methods relying on exact background mappings, the interacting models analyzed in this work represent physically independent scenarios whose observational viability can be assessed directly using current cosmological data.


Within this context, the main goal of the present work is to investigate interacting dark energy models within the CPL framework, considering two commonly studied interaction terms that are linear in the dark sector energy densities, namely $Q \propto H\rho_\text{de}$ and $Q \propto H\rho_\text{c}$. We derive exact analytic expressions for the evolution of dark matter and dark energy in both cases. Unlike previous approaches that rely primarily on numerical methods or impose background-level degeneracies, our formulation provides a fully analytic description of the interacting dark sector, revealing a non-trivial mathematical structure associated with the coupling.


We confront the model with current cosmological observations and investigate how different interaction terms modify the expansion history. In particular, we assess whether such scenarios can produce observable deviations from the $\Lambda$CDM paradigm and potentially contribute to alleviating existing cosmological tensions. Our results provide a unified framework that connects the analytic structure of interacting models with their observational viability, offering new insight into the dynamics of the dark sector.

The paper is organized as follows. In Section \ref{theory} we introduce the interacting dark-sector model and present the cosmological equations governing the evolution of dark matter and dark energy within the CPL parametrization. Furthermore we derive the analytic solutions for the dark-sector energy densities and discuss their mathematical properties. In Section \ref{Cosmo datasamples} we describe the observational datasets and the statistical methodology used to constrain the model parameters. In Section \ref{results} we present the results and analyze the impact of the interaction on the expansion history of the Universe. Finally, in Section \ref{conclusions} we summarize our conclusions and discuss future perspectives.

\section{Theoretical framework}\label{theory}

\subsection{Cosmological equations}

We consider a spatially flat Friedmann–Lemaître–Robertson–Walker (FLRW) universe composed of radiation (r), baryons (b), cold dark matter (CDM), and dark energy (DE). The scale factor satisfies the Friedmann equations \cite{dodelson2021modern}
\begin{eqnarray}
H^2 & = & \frac{8\pi G}{3}\sum_{i}\rho_i, \label{F1} \\
\frac{\ddot{a}}{a} & = & -\frac{4\pi G}{3}\sum_{i}(\rho_i + 3 p_i) \label{f2}.
\end{eqnarray}
Here, $H=\dot{a}/a$ is the Hubble rate, and $\rho_i$ and $p_i$ are the energy density and pressure for each fluid component, where $i=\{\text{r,b,c,de}\}$ denotes radiation, baryons, cold dark matter, and dark energy, respectively. These densities and pressures are linked by the Equation of State (EoS) parameter $w_i=p_i/\rho_i$. For each component: radiation has $w_\text{r} = 1/3$, baryons and cold dark matter are pressureless with $w_\text{b}=w_\text{c} = 0$, and for dark energy, we use the CPL parametrization (\ref{CPL eos}). Next, we consider the possibility that dark matter and dark energy are the only components that interact. Even with this interaction, the total energy density of the dark sector remains conserved, and the energy-momentum tensor for the dark sector, $T_{\nu}^{\mu}=T_{(\text{de})_{\nu}}^{\mu}+T_{(\text{c})_{\nu}}^{\mu}$, satisfies
\begin{equation}
    \nabla_{\mu}T_{\nu}^\mu=\nabla_{\mu}(T_{(\text{de})_{\nu}}^{\mu}+T_{(\text{c})_{\nu}}^{\mu})=0.
\end{equation}
Therefore, if energy is exchanged between DE and DM, as described in \cite{Valiviita:2008iv,Bolotin2015},
\begin{equation}\label{int1}
    \nabla_{\mu}T_{(\text{de})_{\nu}}^{\mu}=-\nabla_{\mu}T_{(\text{c})_{\nu}}^{\mu}=F_\nu,
\end{equation}
where $F_\nu$ is the four-vector that describes the interaction between the dark components.
We can project Eqs. (\ref{int1}) along the four-velocity $u^{\mu}$ to get
\begin{equation}
    u^{\mu}\nabla^{\nu}T_{(\text{c})_{\mu\nu}}=-u^{\mu}F_\mu, \qquad
    u^{\mu}\nabla^{\nu}T_{(\text{de})_{\mu\nu}}=u^{\mu}F_\mu. \label{par}
\end{equation}
For the part orthogonal to the velocity, we use the projector $h_{\beta\mu}=g_{\beta\mu}-u_\beta u_\mu$, obtaining the following:
\begin{equation}
  h^{\mu\beta}\nabla^{\nu}T_{(\text{c})_{\mu\nu}}=-h^{\mu\beta}F_\mu,\quad     h^{\mu\beta}\nabla^{\nu}T_{(\text{de})_{\mu\nu}}=h^{\mu\beta}F_\mu. \label{per}
\end{equation}
The general form of the energy-momentum tensor for a perfect fluid is \cite{carrollspacetime, piattella2018lecture}:
\begin{equation}\label{em}
    T^{\mu\nu}=(\rho+p)u^\mu u^\nu-pg^{\mu\nu}.
\end{equation}
Using Eqs.  (\ref{par}), (\ref{per}) and (\ref{em}), we get the following Euler equations:
\begin{gather}
h^{\mu\beta}\nabla_{\mu}p_\text{c}+(\rho_\text{c}+p_\text{c})u^{\mu}\nabla_{\mu}u^{\beta}=-h^{\mu\beta}F_{\mu}, \label{euler1} \\
h^{\mu\beta}\nabla_{\mu}p_\text{de}+(\rho_\text{de}+p_\text{de})u^{\mu}\nabla_{\mu}u^{\beta}=h^{\mu\beta}F_{\mu}. \label{euler2}
\end{gather}
In the context of a flat FLRW universe, with $u^{\mu}=(1,0,0,0)$ in comoving coordinates, one has
\begin{align}
    \nabla_{\mu}u^{\mu} &= 3H, \\
    u^{\mu}\nabla_{\mu}u^{\nu} &= 0.
\end{align}
Defining $u^{\mu}F_{\mu}=Q$, Eqs.~(\ref{euler1}) and (\ref{euler2}) can be written as
\begin{align}
    \dot{\rho}_\text{c}+3H\rho_\text{c} &= -Q, \label{dot rhoc}\\
    \dot{\rho}_\text{de}+3H(\rho_\text{de}+p_\text{de}) &= Q, \label{dot rho de}
\end{align}
where $Q$ characterizes the rate of energy exchange between dark matter and dark energy. In addition, baryons and radiation are assumed to be independently conserved,
\begin{align}
\dot{\rho}_\text{b}+3H\rho_\text{b}=0, \label{bar} \\
\dot{\rho}_\text{r}+4H\rho_\text{r}=0.  \label{rad}
\end{align}

The sign of Q determines the direction of the energy transfer. With the convention adopted in this work, $Q>0$ corresponds to energy flow from dark matter to dark energy, while $Q<0$ corresponds to the opposite case. At the phenomenological level, the interaction term $Q$ is typically modeled as a function of the energy densities and the Hubble rate. Common choices include terms proportional to $\rho_\text{c}$, $\rho_\text{de}$, or linear combinations thereof, as well as more general forms involving nonlinear couplings, time-dependent functions, or additional cosmological quantities.

Now, we can express the cosmic time derivative in terms of derivatives with respect to redshift:
\[
\frac{d}{dt} = -H(1+z)\frac{d}{dz}.
\]
This allows us to rewrite the system of Eqs. (\ref{dot rhoc})--(\ref{dot rho de}) as
\begin{align}
(1+z)\frac{d\rho_\text{c}}{dz} - 3\rho_\text{c} &= \frac{Q}{H}, \label{dot rhocz} \\
\nonumber \frac{d\rho_\text{de}}{dz} &= \left[\frac{3(1+w_0)}{1+z} + \frac{3w_a z}{(1+z)^2}\right]\rho_\text{de} \\
& \phantom{=} \   - \frac{Q}{H(1+z)}. \label{dot rho dez}
\end{align}
Here, we use the CPL parametrization (\ref{CPL eos}) for the dark energy equation of state $w_\text{de}$. The form of $Q$ affects the structure of these equations and whether analytical solutions are possible. Since such solutions are not always available, we focus on interacting dark energy models that allow analytical expressions for the energy densities. This approach lets us study the cosmological dynamics in detail and compare directly with observational data.

Once the energy densities for the different components are determined, the Friedmann equation (\ref{F1}) becomes
\begin{equation}\label{friedmann}
H^2(z) = \frac{8\pi G}{3} \left[ \rho_\text{r}(z) + \rho_\text{b}(z) + \rho_\text{c}(z) + \rho_\text{de}(z) \right].
\end{equation}
It is useful to express these equations in dimensionless form by defining
\begin{equation}
E(z) \equiv \frac{H(z)}{H_0},
\end{equation}
which leads to the dimensionless Friedmann equation
\begin{equation}\label{Ez_final}
E^2(z) = \Omega_\text{r0}(1+z)^4 + \Omega_\text{b0}(1+z)^3 + \Omega_\text{c}(z) + \Omega_\text{de}(z),
\end{equation}
where the redshift dependent density parameters are defined as $
\Omega_i(z) \equiv \frac{\rho_i(z)}{\rho_{\mathrm{crit},0}}$ where $i=\text{r,b,c,de}$ and $ \rho_{\mathrm{crit},0} = \frac{3H_0^2}{8\pi G}$ . Setting $E(z = 0) = 1$, yields the flatness condition
\begin{equation}
\Omega_\text{de0} = 1 - (\Omega_\text{r0} + \Omega_\text{b0} + \Omega_\text{c0}).
\end{equation}

Now, we introduce the deceleration parameter, which characterizes the acceleration of the cosmic expansion. It is defined as
\begin{equation}
q \equiv -\frac{\ddot{a}}{aH^2}.
\end{equation}
Using the acceleration equation (\ref{f2}), the deceleration parameter provides a direct measure of the competition between matter and dark energy components. For practical purposes, it is convenient to express $q$ in terms of $E(z)$. Using the relation between time and redshift derivatives, one obtains
\begin{equation}
q(z) = -1 + (1+z)\frac{1}{E(z)}\frac{dE(z)}{dz}.
\label{qz}
\end{equation}

Furthermore, the effective EoS parameter $w_\text{eff}$ encodes complementary information about the universe's composition, the evolution of its energy density and the dynamics of its expansion. The general expression for $w_\text{eff}$ depends on the first derivative of $E^2(z)$ with respect to the redshift $z$ as follows:
\begin{eqnarray}
    w_\text{eff}(z)=-1+\frac{1+z}{3E^2}\frac{d E^2}{dz}.\label{weff}
\end{eqnarray}

In the next section, we examine two commonly used interaction models.

\subsection{Model CI: $Q = \beta H \rho_{\textnormal{de}}$}\label{subsec:CI}

If the interaction depends on the dark energy density, the equations take the form
\begin{align}
(1+z)\frac{d\rho_\text{c}}{dz} - 3\rho_\text{c} &= \beta \rho_\text{de}, \\
\frac{d\rho_\text{de}}{dz} &= \bigg[ \frac{3(1+w_0)-\beta}{1+z}  + \frac{3w_a z}{(1+z)^2}\bigg]\rho_\text{de}.
\end{align}

The exact solution for the dark energy density is:
\begin{equation}
\rho_\text{de}(z) = \rho_\text{de,0} (1+z)^{3(1+w_0+w_a)-\beta}
\exp\left(-\frac{3w_a z}{1+z}\right).
\end{equation}
Substituting this result into the dark matter equation gives:
\begin{align}
\rho_\text{c}(z) &= \rho_{c,0}(1+z)^3  + \beta \rho_{de,0} (1+z)^3  \nonumber \\
 &\phantom{=} \ \times \int_0^z (1+\tilde{z})^{3(w_{0} + w_{a})-\beta - 1}  \exp\left( \frac{-3w_{a}\tilde{z}}{1+\tilde{z}} \right) d\tilde{z} .   
\end{align}
This integral can be rewritten using the standard upper incomplete gamma function, defined as \cite{NIST:DLMF}
\[
\Gamma(s,x)=\int_x^\infty t^{\,s-1}e^{-t}\,dt.
\]
As a result, the dark matter density can be written in closed form as
\begin{multline}
\rho_\text{c}(z) = \rho_\text{c,0}(1+z)^3 + \beta \rho_\text{de,0} e^{-3w_a} (3w_a)(-3w_a)^{K-1} \\
\times \left[\Gamma(-K,-3w_a) - \Gamma\left(-K,-\frac{3w_a}{1+z}\right)\right](1+z)^3,
\end{multline}
where
\[
K = 3(w_0+w_a)-\beta.
\]

It is worth noting that the dark energy density has a simple, closed form analytic expression. In contrast, the incomplete gamma function appears only in the dark matter density solution because of the source term from the interaction.

\subsection{Model CII: $Q = \beta H \rho_\mathrm{c}$}\label{subsec:CII}

For an interaction proportional to the dark matter density, the system becomes
\begin{align}
(1+z)\frac{d\rho_\text{c}}{dz} - 3\rho_\text{c} &= \beta \rho_\text{c}, \\
\frac{d\rho_\text{de}}{dz} &= \left[\frac{3(1+w_0)}{1+z} + \frac{3w_a z}{(1+z)^2}\right]\rho_\text{de} - \frac{\beta \rho_\text{c}}{1+z}.
\end{align}

The dark matter density can be solved directly as
\begin{equation}
\rho_\text{c}(z) = \rho_\text{c,0}(1+z)^{3+\beta}.
\end{equation}
The corresponding dark energy density is given by
\begin{multline}
\rho_\text{de}(z) = (1+z)^{3(1+w_0+w_a)}
\exp\left(\frac{-3w_az}{1+z}\right) \\
\times \bigg[\rho_\text{de,0}- \beta \rho_\text{c,0} \int_0^{z} (1+\tilde{z})^{N} \exp \left( \frac{3w_a\tilde{z}}{1+\tilde{z}}\right)  d\tilde{z}  
\bigg],
\end{multline}
where
\[
N = -3(w_0+w_a)+\beta-1.
\] 

Although this solution can be expressed in closed form, analogously to the previous case, in terms of the incomplete gamma function,
\begin{widetext}
\begin{align}
\rho_\text{de}   &=  (1+z)^{3(1+\omega_{0} + \omega_{a})} \exp \left( \frac{-3w_az}{1+z}\right)  \left(  \rho_\text{de,0}  + \beta \rho_\text{c,0} e^{3w_a}(3w_a)^{N+1} \left[ \Gamma\Big( -(N+1),  3 w_a \Big) - \Gamma\left( -(N+1),  \frac{3 w_a}{1+z} \right) \right]  \right),
\end{align}
\end{widetext}
this representation is not optimal from a practical perspective. In particular, both in its standard analytical definition and in numerical implementations, the incomplete gamma function is typically evaluated more reliably for positive values of its second argument. However, in the present context the parameter $w_a$ is generally negative, as will be shown in the Results section, which implies that the arguments of the incomplete gamma function can also become negative. This leads to numerical difficulties associated with the analytic continuation of the function, potentially affecting the stability and accuracy of the computations. For this reason, we prefer to work with the integral representation of the solution, which is numerically more robust and avoids ambiguities related to the evaluation of special functions outside their standard domain. 

\subsection{Effective EoS}

Although the physical equations of state are $w_\text{c}=0$ for cold dark matter and $w_\text{de}(z)$ given by the CPL parametrization \eqref{CPL eos}, it is useful to introduce effective equations of state that allow the interacting system to be recast in a formally conserved form without interaction between the dark fluids.

We define the effective continuity equation
\begin{equation}\label{eq:29}
\dot{\rho}_\text{de} + 3H\left(1+w^{\mathrm{eff}}_\text{de}\right)\rho_\text{de} = 0.
\end{equation}
By comparing Eq.~(\ref{eq:29}) with Eq.~(\ref{dot rho de}), the effective EoS for dark energy is obtained as
\begin{equation}\label{eq:30}
w^{\mathrm{eff}}_\text{de} = w_\text{de}(z) - \frac{Q}{3H\rho_\text{de}}.
\end{equation}
where $w_\text{de}(z)$ corresponds to the CPL parametrization \eqref{CPL eos}.

For the model \textbf{CI}, with $Q=\beta H\rho_\text{de}$, Eq.~(\ref{eq:30}) leads to
\begin{equation}
w^{\mathrm{eff}}_\text{de}(z) = w_\text{de}(z) - \frac{\beta}{3}.
\end{equation}
For the model \textbf{CII}, with $Q=\beta H\rho_\text{c}$, Eq.~(\ref{eq:30}) yields
\begin{equation}
w^{\mathrm{eff}}_\text{de}(z) = w_\text{de}(z)
- \frac{\beta}{3}\frac{\rho_\text{c}(z)}{\rho_\text{de}(z)}.
\label{weffde}
\end{equation}

Therefore, the effective equation of state encodes the impact of the interaction at the background level, while the fundamental physical equation of state remains unchanged.

\subsection{Mathematical structure and viability conditions}

The analytical solutions derived above involve the upper incomplete gamma function, with arguments
\[
x_1 = -3w_a, \qquad x_2 = -\frac{3w_a}{1+z}.
\]

To avoid ambiguities associated with the analytic continuation of the gamma function, we restrict the analysis to the region $w_a \le 0$, ensuring that $x_1$ and $x_2$ remain non-negative for $z \ge 0$.

Additionally, to prevent singular behavior related to the poles of the gamma function, the parameter space is restricted away from non-positive integer values of
\[
K = 3(w_0+w_a)-\beta.
\]

Finally, the physical viability of the model requires
\[
\rho_\text{c}(z) \ge 0, \qquad \rho_\text{de}(z) \ge 0, \qquad H^2(z) > 0,
\]
which guarantees a consistent cosmological evolution.

These conditions define the region of parameter space where the analytical solutions are both mathematically well defined and physically viable. In practice, they are naturally satisfied within the parameter ranges favored by observational constraints, and therefore do not impose additional restrictions on the analysis.

\section{Cosmological data samples}\label{Cosmo datasamples}

We use a combination of cosmological probes to constrain the $\Lambda$CDM, CPL, and interacting models (CI and CII) through a Bayesian Markov Chain Monte Carlo (MCMC) analysis.

\subsection{Observational Hubble Data}

Observational Hubble data (OHD) provide direct constraints on the expansion history via the differential age method based on cosmic chronometers \cite{cosmicchronometers}. These are passively evolving galaxies that allow reliable estimates of the Hubble parameter through
\begin{equation}
H(z) = -\frac{1}{1+z}\frac{dz}{dt}.
\end{equation}
We use $N=34$ measurements in the range $0.07 < z < 1.965$ \cite{HZnewpoint1,hznewpoint2,hznewpoint3}. The corresponding chi-square is
\begin{equation}
\chi^2_{\mathrm{OHD}} = \sum_{i=1}^{N} \left( \frac{H_{\mathrm{th}}(z_i) - H_{\mathrm{obs}}(z_i)}{\sigma_{i,\mathrm{obs}}} \right)^2.
\end{equation}

\subsection{Cosmic Microwave Background}

We include CMB distance priors based on the acoustic scale $l_A$, the shift parameter $R$ \cite{SHIFT_PARAMETER}, and the decoupling redshift $z_\ast$, adopting the covariance matrix from \cite{CMB_DATA}. The chi-square is
\begin{equation}
\chi^2_{\mathrm{CMB}} = X^T \, \mathrm{Cov}^{-1}_{\mathrm{CMB}} \, X,
\end{equation}
where $X$ contains the differences between theoretical and observed quantities. The relevant parameters are
\begin{equation}
l_A = \frac{\pi r(z_\ast)}{r_s(z_\ast)}, \qquad
R = \frac{\sqrt{\Omega_\text{m0}}\,H_0}{c}\,r(z_\ast),
\end{equation}
and $z_\ast$ is computed using the fitting formula of \cite{Hu_1996}. The comoving transverse distance is
\begin{equation}
r(z) = \frac{c}{H_0}\int_0^z \frac{dz'}{E(z')}.
\end{equation}

\subsection{Type Ia Supernovae}

We use the Pantheon+ compilation of 1701 SNIa in the redshift range $0.001 < z < 2.26$ \cite{DATASN}. The chi-square is constructed as
\begin{equation}
\chi^2_{\mathrm{SNIa}} = a + \log\left(\frac{c}{2\pi}\right) - \frac{b^2}{c},
\end{equation}
where $a$, $b$, and $c$ are defined in terms of the covariance matrix and residuals. The luminosity distance is
\begin{equation}
D_L(z) = (1+z)\frac{c}{H_0} \int_0^z \frac{dz'}{E(z')}.
\end{equation}

\subsection{Baryon Acoustic Oscillations}

BAO measurements provide a standard ruler determined by the sound horizon at the drag epoch \cite{Bernal2020}. We include transverse BAO data \cite{DATABAO} and isotropic measurements from DESI DR2 \cite{desi2025}.

The main observables are
\begin{gather}
\nonumber \theta_{\mathrm{BAO}}(z) = \frac{r_d}{(1+z) D_A(z)}, \\
D_V(z) = \left[z D_H(z) D_M^2(z)\right]^{1/3},
\end{gather}
where $D_H(z)=c/H(z)$ is the Hubble distance, $D_M(z)=r(z)$ is the comoving transverse distance, and $D_A(z)$ is the angular diameter distance, defined as
\begin{equation}
D_A(z) = \frac{D_M(z)}{1+z}.
\end{equation}

The sound horizon and drag redshift are computed following \cite{BAOFORMULA}. The chi-square is
\begin{equation}
\chi^2_{\mathrm{BAO}} = \sum_i \frac{\left(O_i^{\mathrm{obs}} - O_i^{\mathrm{th}}(z_i)\right)^2}{\sigma_i^2}.
\end{equation}

All datasets are assumed to be independent, so that the total chi-square (Joint) is given by
\begin{equation}
\chi^2_{\mathrm{Joint}} = \chi^2_{\mathrm{OHD}} + \chi^2_{\mathrm{CMB}} + \chi^2_{\mathrm{SNIa}} + \chi^2_{\mathrm{BAO}}.
\end{equation}

\section{Analyses and results}\label{results}

\subsection{Methodology}

We perform a Bayesian analysis using the MCMC sampler \texttt{emcee}~\cite{mcmchammer}. The analysis is carried out with 400 walkers initialized around the maximum likelihood region, followed by a burn-in phase to ensure convergence. After convergence is achieved, we generate chains of 5000 steps. 

The convergence of the chains is assessed using the autocorrelation time criterion~\cite{Sokal1996MonteCM}. We adopt the following flat priors for the model parameters in both interacting scenarios, summarized in Table \ref{tab:priors}.

\begin{table}[t]
\centering
\begin{tabular}{cc}
\hline
Parameter & Priors \\
\hline
$h$ & $[0.4,\,1.0]$ \\
$\Omega_\text{c}$ & $[0.2,\,0.4]$ \\
$\Omega_\text{b}$ & $[0.01,\,0.06]$ \\
$\beta$ & $[-0.3,\,0.3]$ \\
$w_0$ & $[-3.0,\,-0.5]$ \\
$w_a$ & $[-3.0,\,-0.1]$ \\
\hline
\end{tabular}
\caption{Flat priors adopted for the $\Lambda$CDM, CPL, CI, and CII models.}
\label{tab:priors}
\end{table}

\subsection{Constraints on model parameters}

\begin{figure*}
    \centering
    \includegraphics[width=1.0\linewidth]{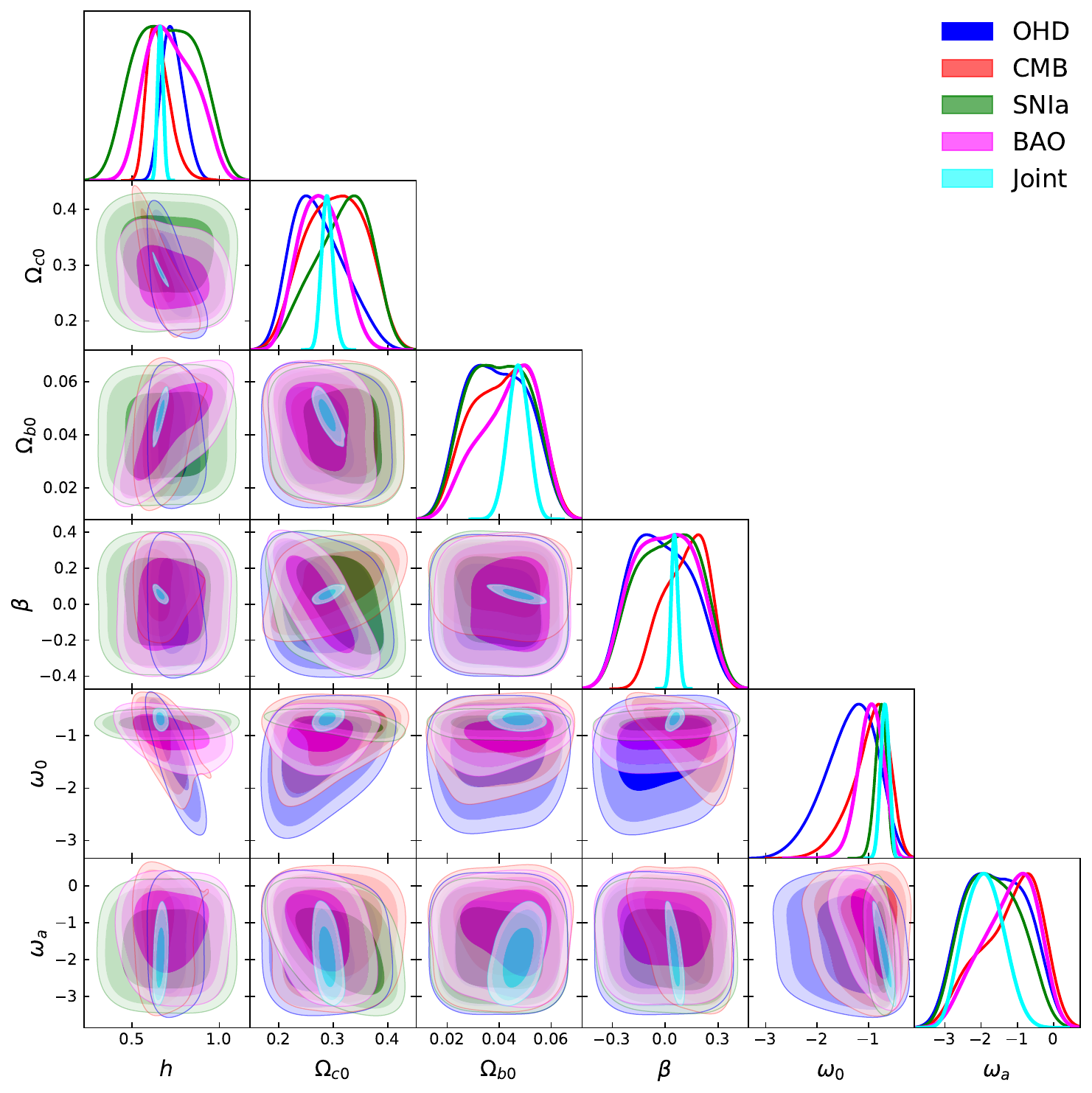}
    \caption{1D posterior distributions and 2D confidence contours at the 1$\sigma$, 2$\sigma$ and 3$\sigma$ (from darker to lighter color bands) for the free parameters characterizing the model CI, using OHD, CMB, SNIa, BAO, and their Joint combination.}
    \label{fig:Contours CI}
\end{figure*}

\begin{figure*}
    \centering
    \includegraphics[width=1.0\linewidth]{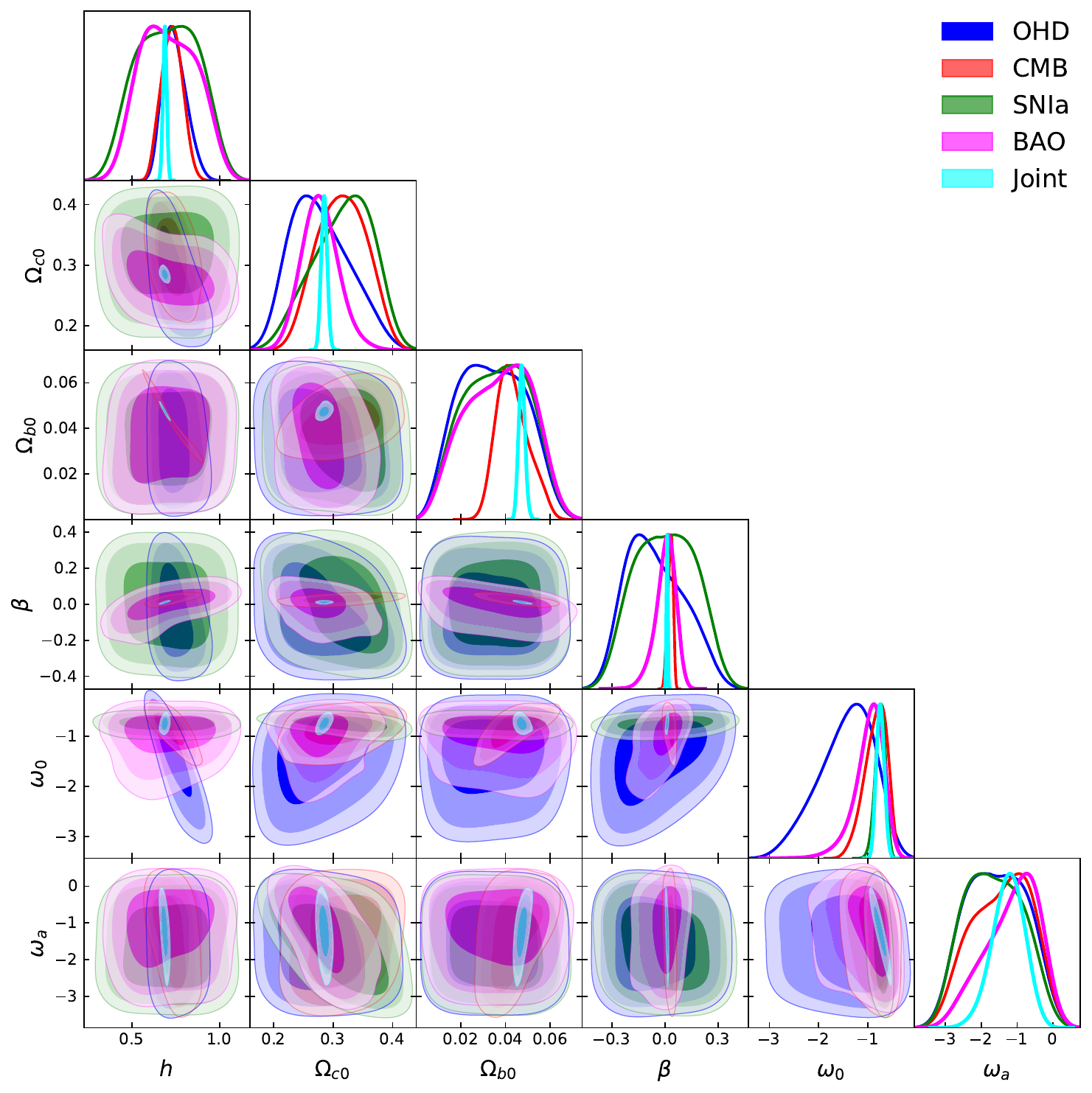}
    \caption{1D posterior distributions and 2D confidence contours to 1$\sigma$, 2$\sigma$ and 3$\sigma$ (from darker to lighter color bands) for the free parameters characterizing the model CII, using OHD, CMB, SNIa, BAO, and their Joint combination.}
    \label{fig:Contours CII}
\end{figure*}

\begin{figure*}
    \centering
    \includegraphics[width=1.0\linewidth]{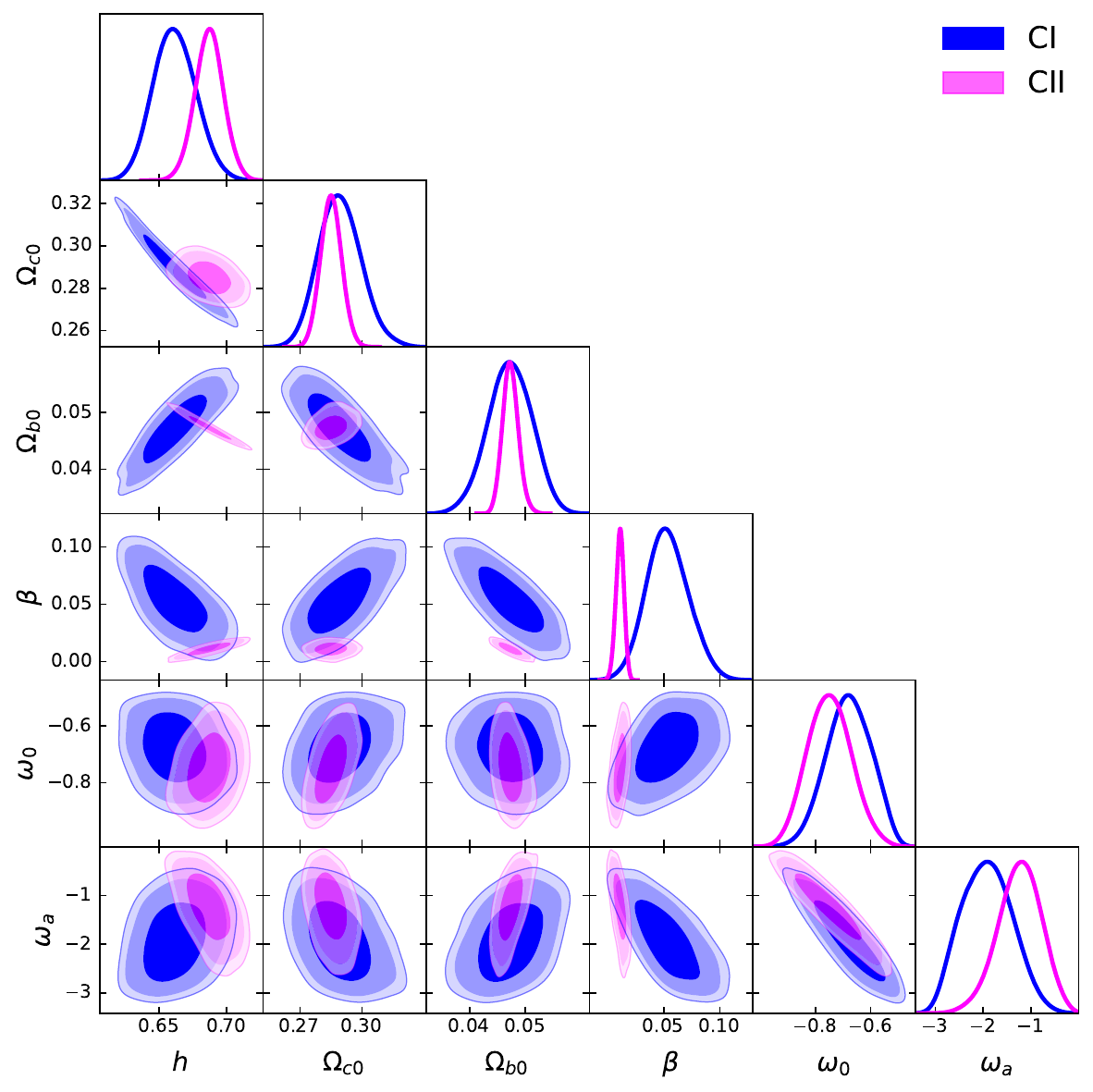}
    \caption{1D posterior distributions and 2D confidence contours to 1$\sigma$, 2$\sigma$ and 3$\sigma$ (from darker to lighter color bands) for the free parameters characterizing the models CI and CII from the Joint analysis.}
    \label{fig:Contours CI-CII}
\end{figure*}

The constraints derived from the Joint analysis of observational Hubble data (OHD), Type Ia supernovae (SNIa), baryon acoustic oscillations (BAO), and cosmic microwave background (CMB) data are summarized in Tables~\ref{tab:CPL}, \ref{tab:CI}, and \ref{tab:CII} for the non-interacting CPL, CI, and CII models, respectively. The corresponding one-dimensional posterior distributions and two-dimensional confidence contours at the 1$\sigma$, 2$\sigma$, and 3$\sigma$ levels are displayed in Figs.~\ref{fig:Contours CI} and \ref{fig:Contours CII} for the CI and CII models. For direct comparison between the two interacting scenarios, Figure~\ref{fig:Contours CI-CII} presents the Joint posterior distributions and confidence contours obtained from the combined datasets.

Figure \ref{fig:Contours CI} shows that, for the CI model, the coupling parameter $\beta$ exhibits
a deviation from zero; however, it remains compatible with non-interacting CPL model within the 1$\sigma$ confidence level. A linear anti-correlation between $\beta$ and $\Omega_\text{c}$ is observed in BAO, 
whereas CMB data exhibit a positive correlation, indicating that the coupling 
parameter $\beta$ plays a non-negligible role at early times. Furthermore, an  increase in the $\beta$ parameter
is offset by a less negative $w_0$. This behavior
indicates that the interaction term can replicate the effect of a dynamical equation
of state, resulting in similar expansion histories.

Among the different datasets, OHD and SNIa constraints are broader, whereas BAO
and CMB provide tighter bounds. The combined analysis significantly reduces the
allowed parameter space, demonstrating that the combination of datasets helps to
break the degeneracy between the interaction parameter and the dark energy equation of state.






Figure \ref{fig:Contours CII} demonstrates that, within the CII model, the coupling parameter $\beta$ is fully consistent with zero at the $3\sigma$ confidence level, thereby maintaining compatibility with the non-interacting CPL scenario. In comparison to the CI case, the contours are more degenerate and less constraining. This outcome indicates that current observations do not impose strong constraints on an interaction proportional to the dark matter density. A mild linear correlation between $\beta$ and both $h$ and $\Omega_\text{c}$ is observed in the BAO and CMB datasets, which suggests that the coupling parameter may still influence early cosmological evolution.

Figure \ref{fig:Contours CI-CII} presents the Joint constraints for both CI and CII models. Both scenarios display a shift toward positive values of $\beta$, although the CII model remains centered around $\beta \simeq 0$. In the CI case, the contours indicate stronger linear correlations between $\beta$ and the dark energy parameters $w_0$ and $w_a$, as well as with $h$, $\Omega_\text{c}$, and $\Omega_\text{b}$. By contrast, the CII model exhibits weaker correlations and larger uncertainties. Nevertheless, the CII scenario demonstrates a partial alleviation of the Hubble tension at the $1\sigma$ level when $\beta$ shifts toward higher values.


The interaction in the CI model introduces significant effects on the cosmological dynamics.
In contrast, the interaction in the CII model remains subdominant.
This distinction indicates that only interactions proportional to the dark energy density produce observable deviations from $\Lambda$CDM within current observational uncertainties.

The contour plots demonstrate that both models are consistent with
$\Lambda$CDM at the $3\sigma$ confidence level. However, the CI model exhibits a
slight preference for a non-zero interaction, whereas the CII model remains consistent with the non-interacting scenario.

\begin{table*}
\centering
{\renewcommand{\arraystretch}{1.2}
{%
\begin{tabular}{|c|c|c|c|c|c|c|c|}
\hline
\multicolumn{8}{|c|}{CPL}\\
\hline
Data set & $h$ & $\Omega_\mathrm{c0}$ & $\Omega_\mathrm{b0}$ & $w_0$ & $w_a$ & $\chi^2$ & $\chi^2_{\text{red}}$ \\
\hline
OHD &
$0.72904^{+0.07100}_{-0.05832}$ &
$0.26622^{+0.05547}_{-0.04275}$ &
$0.03325^{+0.01775}_{-0.01602}$ &
$-1.28379^{+0.45699}_{-0.57020}$ &
$-1.59343^{+0.98436}_{-0.95242}$ &
$27.757$ & $0.957$ \\
\hline
CMB &
$0.71207^{+0.04663}_{-0.05889}$ &
$0.24132^{+0.04553}_{-0.02888}$ &
$0.04407^{+0.00831}_{-0.00524}$ &
$-0.76775^{+0.17365}_{-0.23190}$ &
$-1.23152^{+0.75886}_{-1.00138}$ &
$226.346$ & $75.448$ \\
\hline
SNIa &
$0.71220^{+0.28729}_{-0.31165}$ &
$0.32575^{+0.07376}_{-0.12461}$ &
$0.03492^{+0.02498}_{-0.02484}$ &
$-0.74429^{+0.23338}_{-0.34765}$ &
$-1.70503^{+1.59786}_{-1.29266}$ &
$1756.509$ & $1.036$ \\
\hline
BAO &
$0.66925^{+0.18813}_{-0.12672}$ &
$0.27825^{+0.02911}_{-0.02779}$ &
$0.04344^{+0.01160}_{-0.01885}$ &
$-0.92910^{+0.23078}_{-0.24203}$ &
$-1.09939^{+0.71637}_{-1.01339}$ &
$40.829$ & $2.148$\\
\hline
Joint &
$0.66185^{+0.00564}_{-0.00558}$ &
$0.28396^{+0.00508}_{-0.00484}$ &
$0.05070^{+0.00095}_{-0.00092}$ &
$-0.85616^{+0.06116}_{-0.05205}$ &
$-0.48096^{+0.22889}_{-0.28478}$ &
$1837.529$ & $1.046$ \\
\hline
\end{tabular}
}}
\caption{Best-fit values and corresponding $1\sigma$ errors for the non-interacting CPL model parameters in a flat universe using OHD, CMB, SNIa, BAO, and their Joint combination.}
\label{tab:CPL}
\end{table*}

\begin{table*}
\centering
{\renewcommand{\arraystretch}{1.2}
\resizebox{0.98\textwidth}{!}{%
\begin{tabular}{|c|c|c|c|c|c|c|c|c|}
\hline
\multicolumn{9}{|c|}{CI}\\
\hline
Data set & $h$ & $\Omega_\mathrm{c0}$ & $\Omega_\mathrm{b0}$ & $\beta$ & $w_0$ & $w_a$ & $\chi^2$ & $\chi^2_{\text{red}}$ \\
\hline
OHD &
$0.72752^{+0.07034}_{-0.05755}$ &
$0.26545^{+0.05909}_{-0.04376}$ &
$0.03878^{+0.01412}_{-0.01305}$ &
$-0.03370^{+0.20784}_{-0.18132}$ &
$-1.28648^{+0.46171}_{-0.57274}$ &
$-1.60776^{+0.99806}_{-0.94604}$ &
$27.814$ & $0.993$ \\
\hline
CMB &
$0.64128^{+0.07433}_{-0.05637}$ &
$0.30567^{+0.06059}_{-0.06144}$ &
$0.04261^{+0.01229}_{-0.01530}$ &
$0.13398^{+0.11707}_{-0.16265}$ &
$-0.88974^{+0.28932}_{-0.46083}$ &
$-1.00675^{+0.67383}_{-1.17728}$ &
$369.381$ & $123.127$ \\
\hline
SNIa &
$0.69501^{+0.20523}_{-0.19522}$ &
$0.32344^{+0.04739}_{-0.06454}$ &
$0.03981^{+0.01339}_{-0.01336}$ &
$0.02443^{+0.19008}_{-0.21599}$ &
$-0.75453^{+0.10411}_{-0.11505}$ &
$-1.72352^{+0.90400}_{-0.83623}$ &
$1756.503$ & $1.036$ \\
\hline
BAO &
$0.70673^{+0.17693}_{-0.13198}$ &
$0.27542^{+0.04189}_{-0.04040}$ &
$0.04562^{+0.00992}_{-0.01515}$ &
$0.00101^{+0.19272}_{-0.20245}$ &
$-0.94780^{+0.24069}_{-0.24918}$ &
$-1.10851^{+0.69185}_{-0.95694}$ &
$42.615$ & $2.367$ \\
\hline
Joint &
$0.66093^{+0.04903}_{-0.04119}$ &
$0.28894^{+0.03313}_{-0.02763}$ &
$0.04727^{+0.01074}_{-0.01171}$ &
$0.05252^{+0.05780}_{-0.05279}$ &
$-0.67865^{+0.17506}_{-0.24350}$ &
$-1.93074^{+1.52173}_{-1.05994}$ &
$1828.508$ & $1.041$ \\
\hline
\end{tabular}
}}
\caption{Best-fit values and the corresponding $1\sigma$ errors for the interacting CPL model CI parameters using OHD, CMB, SNIa, BAO, and their Joint combination.}
\label{tab:CI}
\end{table*}

\begin{table*}
\centering
{\renewcommand{\arraystretch}{1.2}
\resizebox{0.98\textwidth}{!}{%
\begin{tabular}{|c|c|c|c|c|c|c|c|c|}
\hline
\multicolumn{9}{|c|}{CII}\\
\hline
Data set & $h$ & $\Omega_\mathrm{c0}$ & $\Omega_\mathrm{b0}$ & $\beta$ & $w_0$ & $w_a$ & $\chi^2$ & $\chi^2_{\text{red}}$ \\
\hline
OHD &
$0.73493^{+0.07676}_{-0.06028}$ &
$0.27095^{+0.06177}_{-0.04663}$ &
$0.03388^{+0.01736}_{-0.01639}$ &
$-0.08084^{+0.21496}_{-0.15370}$ &
$-1.35983^{+0.50978}_{-0.66847}$ &
$-1.57581^{+0.98084}_{-0.96421}$ &
$27.939$ & $0.997$ \\
\hline
CMB &
$0.70796^{+0.19274}_{-0.20723}$ &
$0.32299^{+0.04719}_{-0.06104}$ &
$0.03661^{+0.01588}_{-0.01758}$ &
$0.00291^{+0.19109}_{-0.19767}$ &
$-0.74228^{+0.09755}_{-0.10644}$ &
$-1.66929^{+0.94882}_{-0.89742}$ &
$2.678$ & $117.435$ \\
\hline
SNIa &
$0.72715^{+0.06201}_{-0.06539}$ &
$0.31521^{+0.04430}_{-0.04569}$ &
$0.04230^{+0.00873}_{-0.00635}$ &
$0.03026^{+0.01443}_{-0.01782}$ &
$-0.81264^{+0.21233}_{-0.26477}$ &
$-1.33147^{+0.82122}_{-1.05936}$ &
$1756.521$ & $1.0362$ \\
\hline
BAO &
$0.70175^{+0.19395}_{-0.16409}$ &
$0.27653^{+0.03119}_{-0.02861}$ &
$0.03829^{+0.01527}_{-0.01857}$ &
$0.00823^{+0.04783}_{-0.06165}$ &
$-0.92485^{+0.23618}_{-0.32785}$ &
$-0.99121^{+0.64789}_{-0.97232}$ &
$74.995$ & $2.678$ \\
\hline
Joint &
$0.68718^{+0.03103}_{-0.03316}$ &
$0.28499^{+0.01516}_{-0.01456}$ &
$0.04730^{+0.00502}_{-0.00373}$ &
$0.01094^{+0.00999}_{-0.01196}$ &
$-0.74925^{+0.23787}_{-0.20768}$ &
$-1.22132^{+1.07392}_{-1.47160}$ & 
$1835.291$ & $1.045$ \\
\hline
\end{tabular}
}}
\caption{Best-fit values and the corresponding $1\sigma$ errors for the interacting CPL model CII parameters using OHD, CMB, SNIa, BAO, and their Joint combination.}
\label{tab:CII}
\end{table*}

\subsection{Model comparison}

\label{comparison}
After obtaining the constraints from the different cosmological data, we compare the different DE models using the Akaike information criterion (AIC, \cite{1974ITAC...19..716A}) and Bayesian information criterion \cite{BIC1974} defined as:
\begin{equation}
\text{AIC} = \chi^2_{\text{min}} + 2\alpha,
\end{equation}
\begin{equation}
\text{BIC} = \chi^2_{\text{min}} + \alpha \ln N,
\end{equation}
where $\chi^2_{\text{min}}$ is the chi-square obtained from the best fit of the parameters, $\alpha$ is the number of parameters, and $N$ is the number of data points used in the fit. The methodology consists of computing the difference between the value of each information criterion and that of a reference model, defined as the one with the minimum AIC or BIC. Specifically, $\Delta \text{AIC} = \text{AIC}i - \text{AIC}_{\text{min}}$ provides insight into the relative support of the models, whilst $\Delta \text{BIC} = \text{BIC}i - \text{BIC}_{\text{min}}$ quantifies the strength of evidence against a given model, as summarized in Table \ref{table4}.



The model comparison based on information criteria reveals a tension between goodness of fit and model simplicity as shown in Table \ref{tab:AICBIC}. According to the AIC, the interacting model CI provides the lowest AIC value, indicating that the inclusion of an interaction term can improve the likelihood. However, the BIC, which more strongly penalizes the number of free parameters, favors the $\Lambda$CDM model.

In contrast, the CII model is strongly disfavored by both AIC and BIC, providing clear and robust evidence against an interaction proportional to the dark matter density. A similar behavior is observed for the CPL parametrization, which, although more flexible than $\Lambda$CDM, is not statistically preferred by the data.

In general, interacting models, particularly CI, offer a better fit to the data according to the AIC and provide a richer phenomenology, including a dynamical equation of state and possible future transitions. However, this preference is not decisive once the penalty for additional parameters is considered, as reflected by the BIC criteria.

\begin{table}
\centering
\begin{tabular}{|c|c|}
\hline
$\Delta$AIC & Empirical support for model $i$ \\
\hline
0 - 2 & Substantial \\
4 - 7 & Considerably less \\
$>$ 10 & Essentially none \\
\hline
$\Delta$BIC & Evidence against model $i$ \\
\hline
0 - 2 & Not worth more than a bare mention \\
2 - 6 & Positive \\
6 - 10 & Strong \\
$>$ 10 & Very strong \\
\hline
\end{tabular}
\caption{Reference values for the $\Delta$AIC and $\Delta$BIC criteria. Note that smaller values of $\Delta$AIC and $\Delta$BIC indicate a preference for the model.}\label{table4}
\end{table}

\subsection{Cosmological evolution}

\begin{figure*}[t]
  \centering
  \begin{minipage}[b]{0.49\textwidth}
    \centering
    \includegraphics[width=\linewidth]{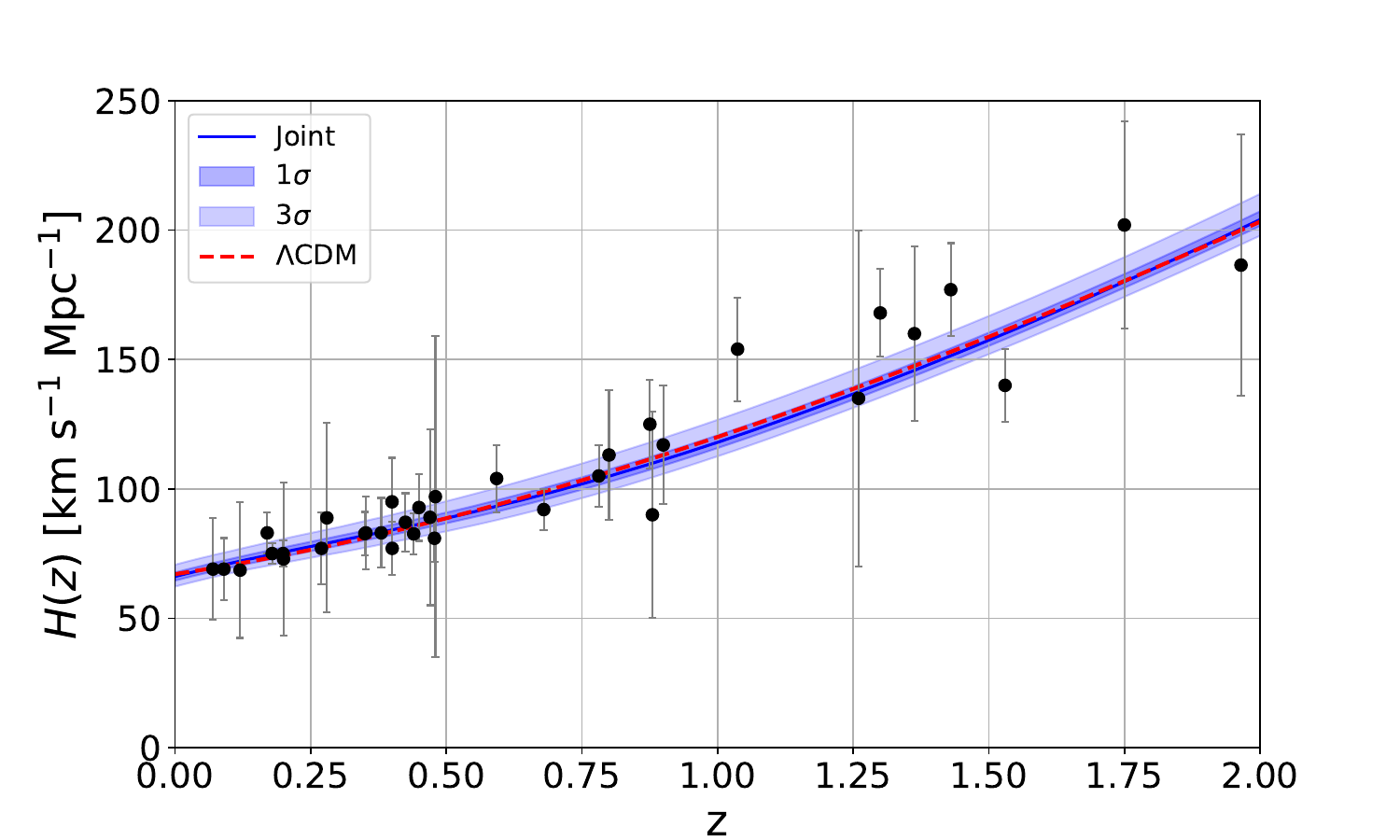}
  \end{minipage}\hfill
  \begin{minipage}[b]{0.49\textwidth}
    \centering
    \includegraphics[width=\linewidth]{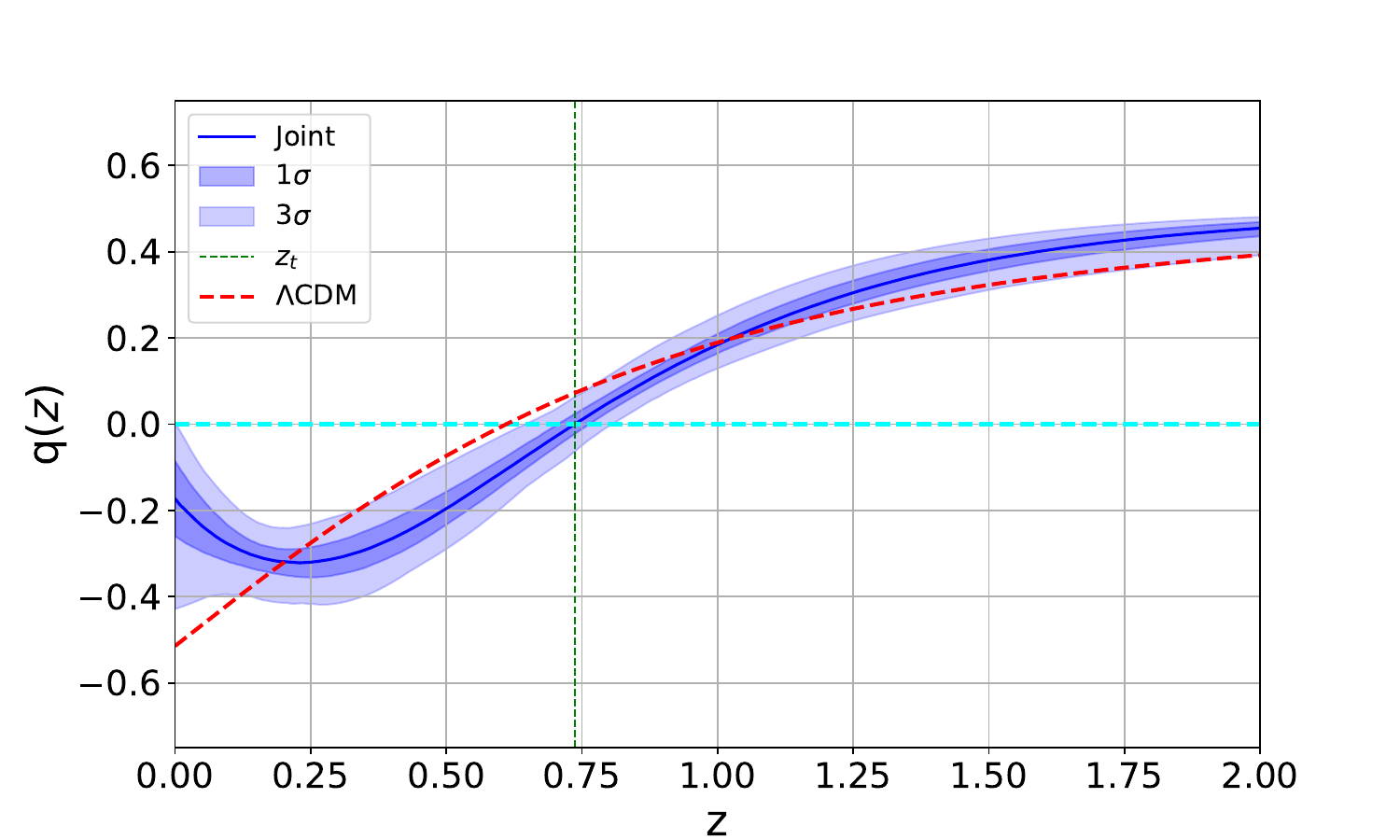}
  \end{minipage}
  \caption{Left panel: Reconstruction of the Hubble rate $H(z)$ with the respective OHD data points and their errors. Right panel: Reconstruction of the deceleration parameter $q(z)$. For both panels the shaded regions represent the $1\sigma$ (darker) and $3\sigma$ (lighter) error bands from the Joint analysis for the CI.}
  \label{fig:HZ CI}
\end{figure*}

\begin{figure*}[t]
  \centering
  \begin{minipage}[b]{0.49\textwidth}
    \centering
    \includegraphics[width=\linewidth]{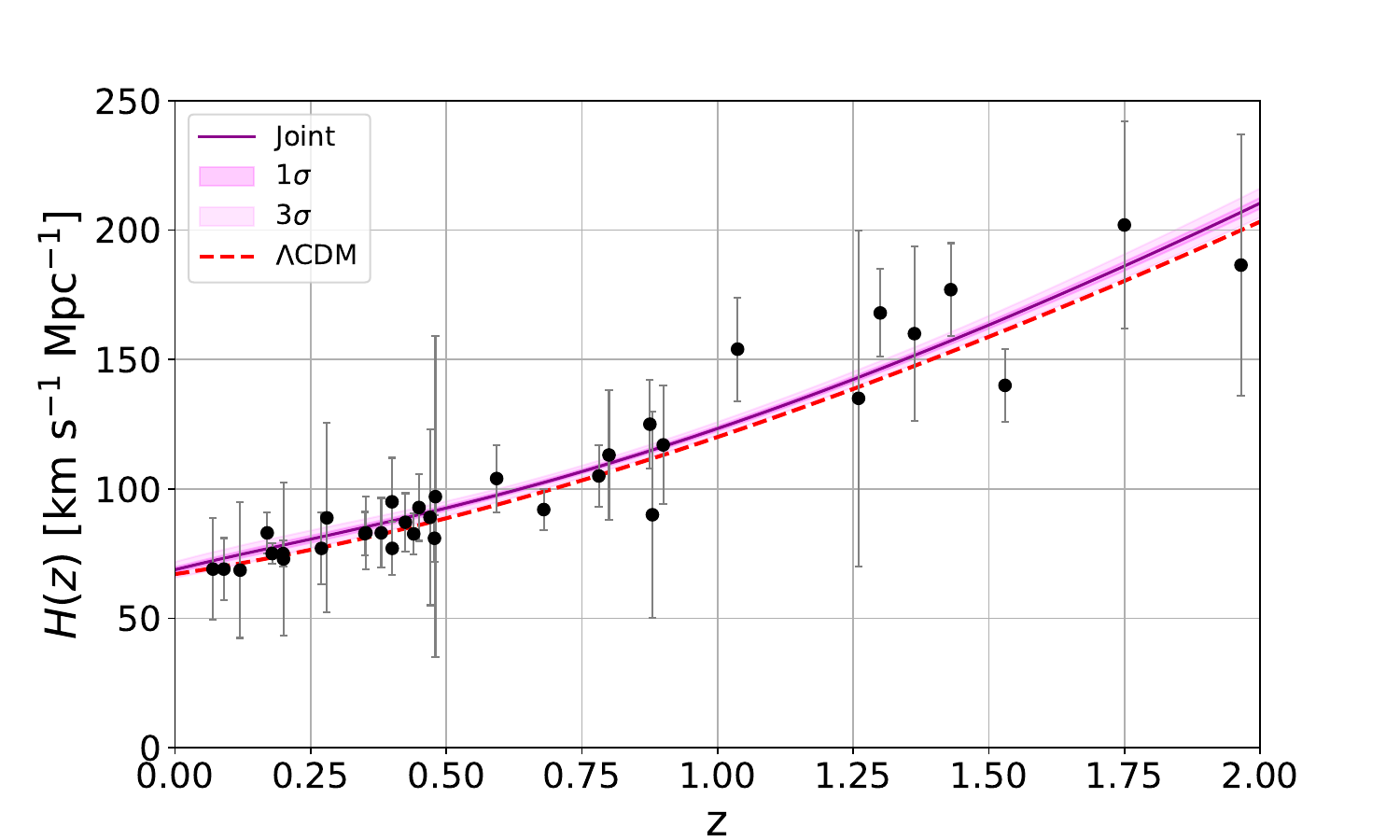}
  \end{minipage}\hfill
  \begin{minipage}[b]{0.49\textwidth}
    \centering
    \includegraphics[width=\linewidth]{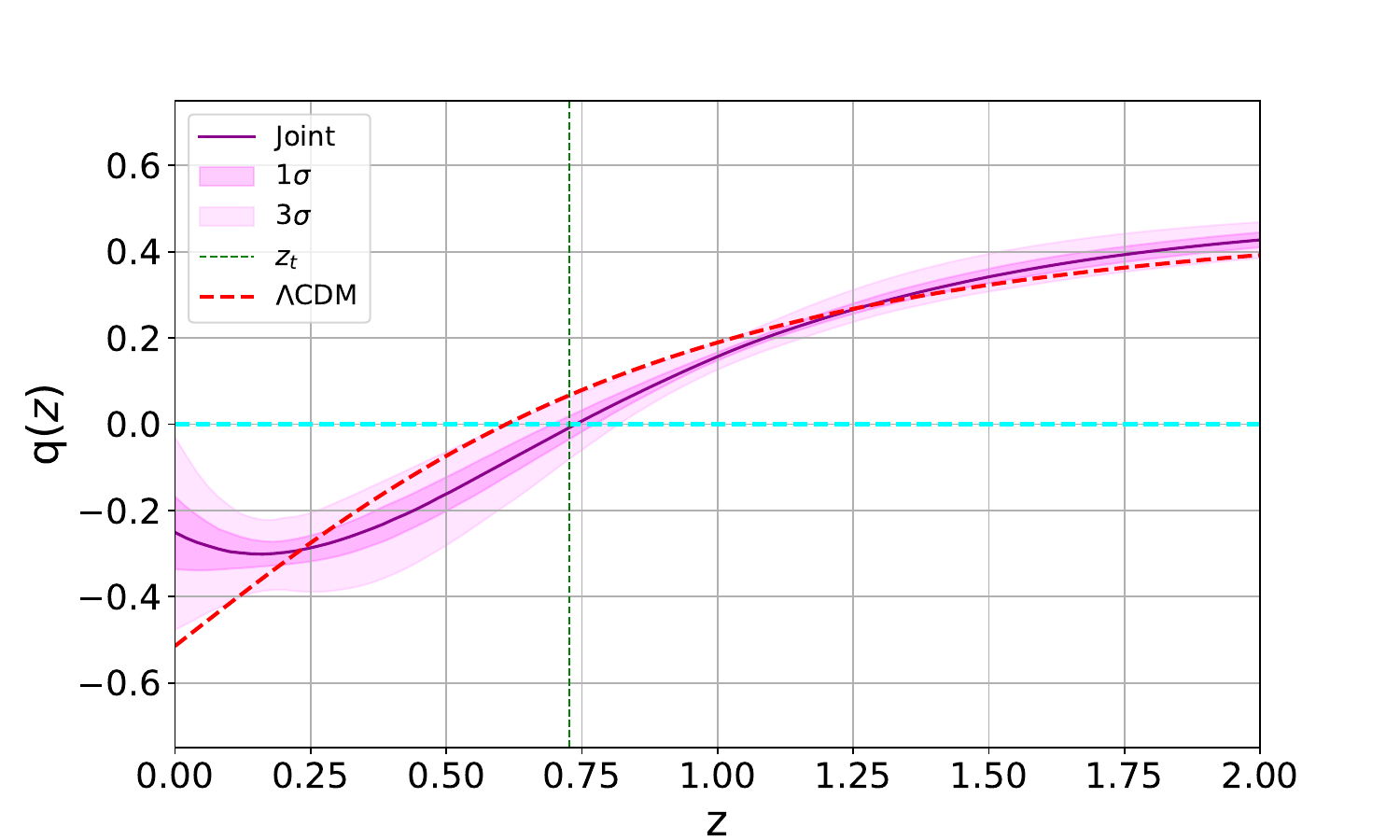}
  \end{minipage}
  \caption{Left panel: Reconstruction of the Hubble rate $H(z)$ with the respective OHD data points and their errors. Right panel: Reconstruction of the deceleration parameter $q(z)$. For both panels the shaded regions represent the $1\sigma$ (darker) and $3\sigma$ (lighter) error bands from the Joint analysis for the CII. }
  \label{fig:HZ CII}
\end{figure*}

\begin{figure*}[t]
  \centering
  \begin{minipage}[b]{0.49\textwidth}
    \centering
    \includegraphics[width=\linewidth]{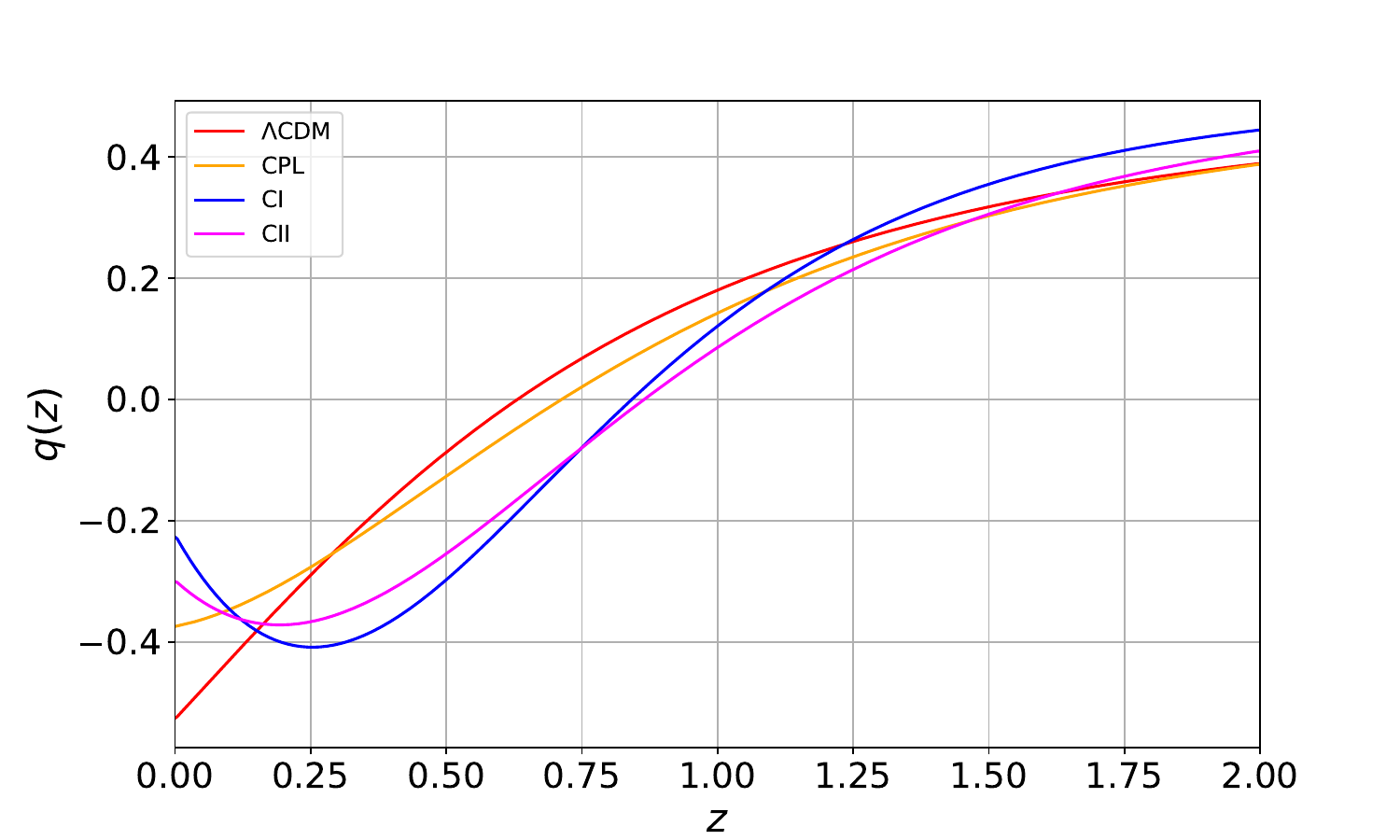}
  \end{minipage}\hfill
  \begin{minipage}[b]{0.49\textwidth}
    \centering
    \includegraphics[width=\linewidth]{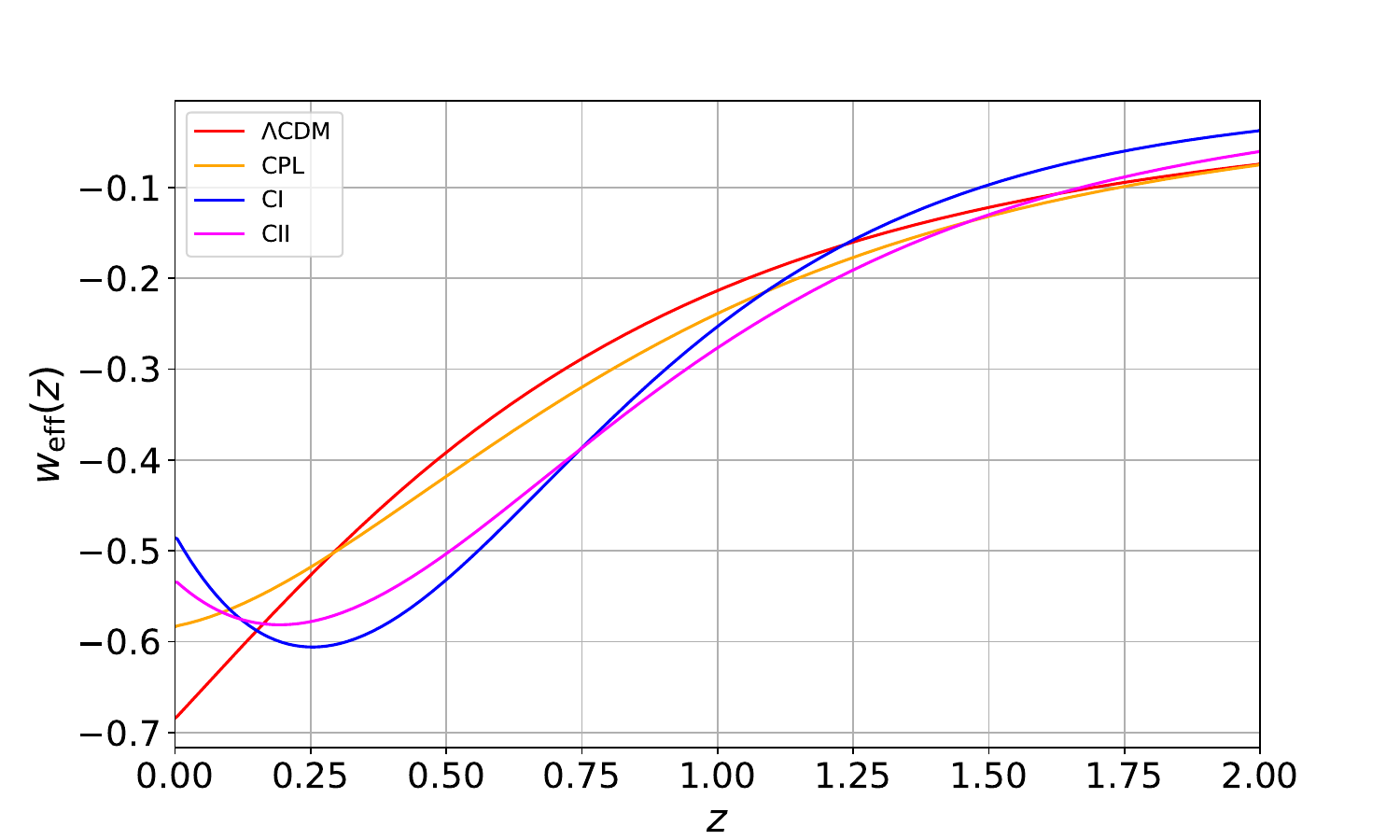}
  \end{minipage}
  \caption{Left panel: Reconstruction of the deceleration parameter $q(z)$ of the models CPL, CI and CII.
  Right panel: Reconstruction of the effective EoS for dark energy $w_\mathrm{eff}$ of the models $\Lambda$CDM, CPL, interacting models CI and CII. Both panels were obtained using the best-fit parameters from the Joint analysis.}
  \label{fig:q-w}
\end{figure*}

\begin{figure}
    \centering
    \includegraphics[width=0.99\linewidth]{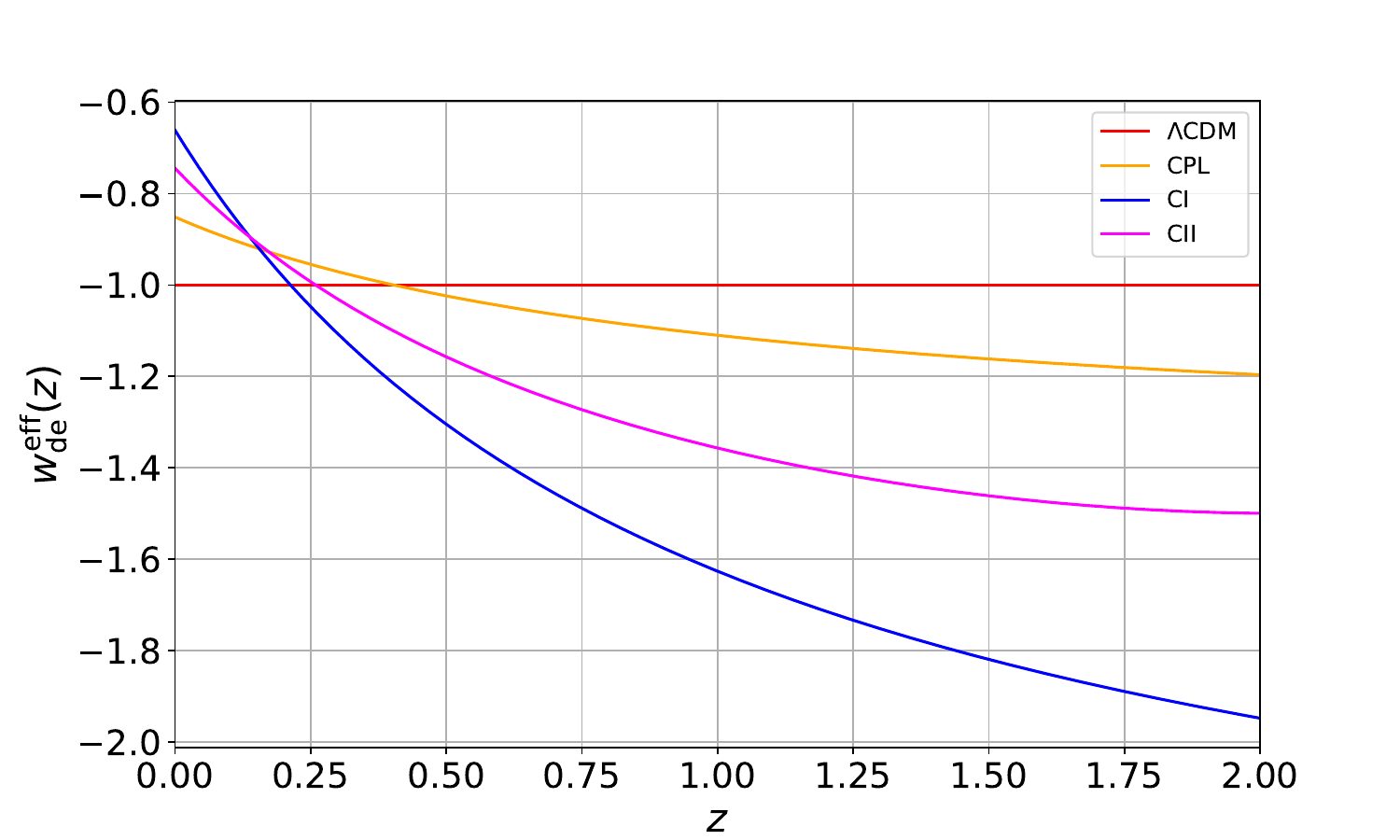}
    \caption{
Redshift evolution of the effective dark energy EoS $w^{\mathrm{eff}}_\text{de}(z)$ for $\Lambda$CDM, CPL, and the interacting models CI and CII, using best-fit parameters from the Joint analysis.
}
    \label{fig:weff-de}
\end{figure}

With the best-fit values obtained from the Joint analysis, we can plot relevant quantities that describe the cosmic evolution at the background level and compare them with those from the $\Lambda$CDM and non-interacting CPL models. Figures \ref{fig:HZ CI}-\ref{fig:weff-de} illustrate the impact of dark-sector interactions on the cosmological evolution. Both models remain consistent with current observational constraints; however, they exhibit systematic deviations from both $\Lambda$CDM and the non-interacting CPL model. 

The evolution of the Hubble rate $H(z)$ for models CI and CII is presented in left panels of Figs. \ref{fig:HZ CI} and \ref{fig:HZ CII}, respectively. Both models are able to reproduce the observed expansion history over the full redshift range within the $1\sigma$ uncertainties compared to $\Lambda$CDM. The deceleration parameter as a function of redshift is presented in right panels of Figs. \ref{fig:HZ CI} and \ref{fig:HZ CII} for models CI and CII, respectively. In all scenarios, the Universe evolves from a matter-dominated phase to a late-time accelerated regime. However, the transition redshift $z_t$, defined by $q(z_t)=0$, is sensitive to the presence of interaction.

At the present epoch, the CI model yields $H_0 = 66.06^{+1.58}_{-1.49}\,\mathrm{km\,s^{-1}\,Mpc^{-1}}$ and $q_0 = -0.173^{+0.086}_{-0.087}$, while the CII model gives $H_0 = 68.73^{+1.02}_{-1.09}\,\mathrm{km\,s^{-1}\,Mpc^{-1}}$ and $q_0 = -0.251^{+0.084}_{-0.085}$. We find that both interacting scenarios remain consistent with the Planck constraint on $H_0$ \cite{Planck:2018vyg} within $1\sigma$, with the CI model slightly favoring lower values, while the CII model favors slightly higher values for the Hubble constant $H_0$ and $q_0$ respectively. In contrast, the inferred present-day deceleration parameter in both models is significantly less consistent than the Planck derived value $q_0 \simeq -0.528$ \cite{Planck:2018vyg}, indicating a weaker current acceleration in the interacting scenarios compared to $\Lambda$CDM.



 In the interacting CI model, a positive coupling can delay the onset of cosmic acceleration compared to the non-interacting CPL model. This effect demonstrates the altered evolution of the dark sector resulting from the interaction. The coupling modifies the relative evolution of dark matter and dark energy densities, thereby altering the background expansion history and shifting the transition redshift relative to the non-interacting CPL scenario. As a result, the transition redshift shifts to lower values, $z_t=0.727^{+0.040}_{-0.040}$. The epoch of maximal acceleration for CI is defined by $z_{\min} = 0.155^{+0.079}_{-0.125}$ and $q(z_{\min}) = -0.313^{+0.032}_{-0.039}$.
In contrast, for the interacting model CII, the coupling parameter is consistent with zero within observational uncertainties, indicating a background evolution that closely matches the non-interacting CPL scenario. Therefore, both the transition redshift and the profile of $q(z)$ are only marginally influenced. Specifically, the minimum of the deceleration parameter for CII takes place at
$z_{\min} = 0.228^{+0.042}_{-0.066}$
and $q(z_{\min}) = -0.328^{+0.030}_{-0.032}$,
which indicates a slightly earlier yet comparably deep acceleration phase relative to the CI case.
In general, although the CI scenario accounts for a more significant deviation of the expansion history in comparison to $\Lambda$CDM and non-interacting CPL scenarios, the CII model becomes indistinguishable from the standard CPL case at the background level.

The evolution of the effective equation of state $w_{\rm eff}(z)$, shown in right panel of Fig. \ref{fig:q-w}, gives further clarification on the impact of interaction. The non-interacting CPL model shows mild redshift dependence, having values close to $-1$ at low redshift, but allowing for moderate deviations at earlier epochs. Conversely, the interacting CI model exhibits a markedly enhanced evolution of $w_{\rm eff}(z)$, especially at intermediate redshifts. The positive coupling $\beta>0$ modifies the dynamics of dark energy, leading to a more rapid evolution of the equation of state than in the non-interacting scenario. This outcome highlights the degeneracy between interaction and dynamical dark energy, as the energy transfer alters the redshift dependence of the dark sector.

In contrast, the CII model remains close to the CPL prediction, exhibiting only minor deviations within the allowed parameter space. This consistency arises because the coupling parameter is statistically compatible with zero, and in this scenario, the interaction influences the effective equation of state only indirectly through the ratio $\rho_\text{c}/\rho_\text{de}$.

In summary, the analysis of cosmic evolution indicates that interacting dark energy models can produce observable deviations from $\Lambda$CDM, especially when the interaction is proportional to the dark energy density. These deviations appear as shifts in the transition redshift and changes in the evolution of the effective equation of state. Nevertheless, current observational data still favor models that closely align with the standard cosmological scenario, providing only limited evidence for interaction.

\begin{table}\label{table3}
\centering
\resizebox{0.98\linewidth}{!}{%
\begin{tabular}{|c|c|c|c|c|c|c|c|c|}
\hline
\textbf{Data set} & \multicolumn{2}{|c|}{$\Lambda$CDM} & \multicolumn{2}{|c|}{CPL} & \multicolumn{2}{|c|}{CI} & \multicolumn{2}{|c|}{CII} \\
\hline
 & $\Delta$AIC & $\Delta$BIC & $\Delta$AIC & $\Delta$BIC & $\Delta$AIC & $\Delta$BIC & $\Delta$AIC & $\Delta$BIC \\
\hline
Joint & $6.287$ & $0.0$ & $7.020$ & $16.155$ & $0.0$ & $17.2356$ & $6.6781$ & $10.455$ \\
\hline
\end{tabular}}
\caption{Differences in AIC and BIC values with respect to the minimum among the listed models, computed using the chi-square value from the Joint analysis.}
\label{tab:AICBIC}
\end{table}

\subsection{Physical interpretation and future evolution}

Expanding on the previously discussed background evolution, we present a physical interpretation of the best-fit parameters and analyze their implications for the Universe's past and future evolution.

The best-fit parameters derived for the interacting model CI,
\[
w_0 \approx -0.68, \qquad w_a \approx -1.83,
\]
suggest the presence of a highly dynamical dark energy component that evolves substantially with redshift.

At high redshift ($z \to \infty$), the CPL parametrization approaches
\[
w_\text{de}(z) \to w_0 + w_a,
\]
which, for the best-fit values, yields
\[
w_\text{de}(z \to \infty) \approx -2.5.
\]
This result corresponds to a phantom-like behavior of the best-fit CPL equation of state at the background level. Since the interaction is absorbed into an effective conserved description, this behavior does not necessarily imply the existence of a fundamental phantom dark-energy field. In contrast, at present, the equation of state approaches a quintessence-like behavior ($w_0>-1$), indicating an effective transition across the phantom divide ($w_\text{de}=-1$) in the background evolution.

A positive coupling $\beta>0$ indicates a transfer of energy
from dark matter to dark energy. The interaction modifies the relative evolution of the
dark-sector components and contributes to the observed
departure from the non-interacting CPL background
evolution. This energy exchange alters the expansion history and
modifies the timing of the transition to accelerated
expansion.

Extending the analysis to negative redshift ($z<0$) offers insight into the future evolution of the Universe. In this regime, the CPL parametrization diverges formally as $z \rightarrow -1$. Therefore, although the parametrization remains reliable at low and intermediate redshifts, extrapolation to the far future must be interpreted with caution \cite{Chevallier:2000qy,Linder:2002et}.

Within the range of validity of the CPL parametrization, the condition $w_a<0$ implies that $w_\text{de}(z)$ increases toward positive values in the future. As a result, the effective equation of state of the whole cosmic fluid may eventually exceed the threshold for accelerated expansion, $w_{\rm eff} < -1/3$. Consequently, the Universe may transition from the current accelerated phase to a future decelerated expansion regime. This behavior can also be understood from the evolution  of the deceleration parameter $q(z)$, which may display two distinct zeros: one at $z_t>0$, marking the onset of the current accelerated expansion, and another at $z_f<0$, indicating a future transition to deceleration. In the analyzed interacting scenarios, both models predict a future transition, though at slightly different epochs. The CI model yields $z_f = -0.316^{+0.007}_{-0.007}$, while the CII model gives $z_f = -0.251^{+0.118}_{-0.187}$.  This behavior can be interpreted as follows. The future transition to deceleration is primarily determined by the reconstructed CPL dynamics, characterized by $w_a<0$. Both interacting models predict a future decelerated phase. 

Therefore, the future transition is not exclusively associated with the interaction itself. Instead, the interaction modifies the detailed evolution of the dark sector and alters the specific epoch at which the transition occurs. This interpretation aligns with the reconstructed evolution of $q(z)$ and $w_{\rm eff}(z)$ shown in Fig.~\ref{fig:q-w}.

Taken together, these findings indicate that the reconstructed CPL dynamics naturally allow for a transient phase of cosmic acceleration, while dark-sector interactions modify its detailed evolution and the precise transition epoch. This result implies that the present accelerated expansion may not be a permanent feature of the Universe. Similar transient acceleration scenarios have been examined in previous studies \cite{Barrow2000,Vargas2011,Shahalam2015,Hu2015,Magana2014,Zhang2018,Bolotin2020,Escobal:2023,Fortunato:2024}. Although this behavior involves extrapolating the CPL parametrization beyond its validated domain, it highlights the intricate phenomenology arising from the interplay between dark energy dynamics and interactions.

The evolution of the effective dark energy equation of state 
$w^{\mathrm{eff}}_\text{de}(z)$ (\ref{weffde}) also reveals a phantom-like behavior, with $w^{\mathrm{eff}}_\text{de}(z) < -1$ across a range of redshifts, as shown in Fig.~\ref{fig:weff-de}. Notably, even without interaction, the CPL parametrization can yield $w_\text{de}(z) < -1$ at early times for sufficiently negative $w_a$ \cite{Chevallier:2000qy,Linder:2002et}. Thus, the emergence of an effective phantom behavior is not unique to interacting models. Instead, the interaction modifies the detailed redshift evolution of the effective equation of state and the associated background dynamics.

In interacting scenarios, this behavior acquires a distinct physical interpretation. The effective equation of state reflects energy transfer between dark energy and dark matter, and therefore does not represent the intrinsic equation of state of a fundamental fluid. In this context, the condition $w^{\mathrm{eff}}_\text{de} < -1$ should be regarded as an effective description rather than a direct violation of the null energy condition at the level of the total energy-momentum tensor \cite{Bolotin2015,Wang2016}.

Specifically, the interaction can enhance or shift the phantom-like behavior by altering the redshift evolution of the dark energy density. This result highlights the degeneracy between dynamical dark energy and interactions within the dark sector, indicating that observational evidence of $w_\text{de}<-1$ does not uniquely imply the presence of fundamental phantom fields \cite{Guedezounme:2025wav}. Recent analyses in the DESI era and related datasets have further emphasized this degeneracy and the viability of interacting scenarios in explaining deviations from $w_\text{de}=-1$ \cite{DESI:2024aqx,Malekjani:2024bgi}.


\subsection{Perturbative stability considerations}

Although the present work focuses on the background evolution, it is important to briefly comment on the perturbative stability of interacting dark energy models. It is well known that certain choices of the interaction term $Q$ can lead to instabilities at early times or on large scales \cite{Valiviita:2008iv,He:2008si}. The stability of perturbations depends on both the form of the interaction and the behavior of the dark energy equation of state.

Consistent frameworks, such as parametrized post-Friedmann (PPF) approaches and covariant formulations, have been developed to ensure a stable evolution of perturbations in interacting dark energy scenarios \cite{Li:2014eha}. More recent analyses have further explored these issues in the context of high-precision cosmological data, confirming the importance of perturbative stability as a viability criterion for interacting models \cite{Yang:2018euj,Pan:2019gop}. 

In the present analysis, the coupling parameter is constrained to small values, $|\beta| \ll 1$, suggesting that the models lie within a perturbatively viable region of parameter space.

\section{Conclusions}\label{conclusions}

This study investigated interacting dark energy models within the Chevallier-Polarski-Linder (CPL) parametrization, focusing on their analytic structure and observational viability. Two widely studied interaction terms, $Q = \beta H \rho_\text{de}$ (CI) and $Q = \beta H \rho_\text{c}$ (CII), are examined, and exact analytic solutions for the evolution of the dark sector are derived. These solutions reveal a non-trivial mathematical structure involving incomplete gamma functions, a feature often lost in purely numerical analyses.

From an observational standpoint, we constrained the models through a Joint analysis of observational Hubble data (OHD), Type Ia supernovae (SNIa), baryon acoustic oscillations (BAO), and cosmic microwave background (CMB) data. Our findings indicate that the CI model yields a modestly improved fit relative to the non-interacting CPL scenario according to the Akaike Information Criterion (AIC), whereas the Bayesian Information Criterion (BIC) continues to favor the simpler $\Lambda$CDM model. Conversely, the CII model is strongly disfavored, suggesting that interactions proportional to the dark matter density are not supported by current observations.

Regarding cosmological evolution, we find that interactions proportional to the dark energy density can generate observable deviations from $\Lambda$CDM, such as a delayed onset of cosmic acceleration and a modification on the evolution of the effective equation of state. Specifically, the preferred CI scenario demonstrates a transition from an early-time effective phantom regime to a late-time quintessence-like behavior. These results emphasize the degeneracy between dynamical dark energy and dark sector interactions, indicating that observed deviations from $w=-1$ do not necessarily imply the existence of fundamental phantom fields.

Our analysis further suggests the possibility of a future transition from accelerated to decelerated expansion. Because both the CI and CII scenarios exhibit similar behavior, the future transition to deceleration appears to be primarily driven by the reconstructed CPL dynamics rather than by the interaction itself. The interaction mainly affects the detailed evolution of the dark sector and the precise transition epoch. However, this result should be interpreted with caution, given the limitations of the CPL parametrization as $z$ approaches $-1$. 

In summary, interacting dark energy models within the CPL framework represent a viable and adaptable extension of the standard cosmological model, offering more complex phenomenology while remaining consistent with current observations. Nevertheless, statistical evidence for interaction remains limited. Future research incorporating perturbation-level analyses and large-scale structure observables will be crucial to break the degeneracy with dynamical dark energy and to comprehensively evaluate the viability of these scenarios.

\section*{Acknowledgements}

Nelson Videla acknowledges support from ANID--FONDE\-CYT Grant No.~1220065. Gerald Neumann acknowledges support from Dirección de Postgrado of Universidad Técnica Federico Santa María through their grants 'Exención de Arancel' (Tuition Waiver) and 'Apoyo Financiero' (Financial Support) under CONVENIO DE BECA DE DOCTORADO N° 062 / 2026. Dorian Araya acknowledges support from ANID, Beca de Magíster Nacional, Año Académico 2026, Grant No.~22262426.

\appendix
\section{Detailed derivations}

In this appendix we present the derivation of the analytic solutions for the interacting dark sector models considered in Section 2.

\subsection{General system of equations}

We consider the conservation equations
\begin{align}
\dot{\rho}_\text{c} + 3H\rho_\text{c} &= -Q, \\
\dot{\rho}_\text{de} + 3H\rho_\text{de}(1+w_\text{de}) &= Q,
\end{align}
together with the CPL parametrization
\begin{equation}
w_\text{de}(z) = w_0 + w_a \frac{z}{1+z}.
\end{equation}

Using the relation
\[
\frac{d}{dt} = -H(1+z)\frac{d}{dz},
\]
the system becomes
\begin{align}
(1+z)\frac{d\rho_\text{c}}{dz} - 3\rho_\text{c} &= \frac{Q}{H}, \\
\frac{d\rho_\text{de}}{dz} &= \left[\frac{3(1+w_0)}{1+z} + \frac{3w_a z}{(1+z)^2}\right]\rho_\text{de} \nonumber \\
& \phantom{=} \ - \frac{Q}{H(1+z)}.
\end{align}

\subsection{Model CI: $Q = \beta H \rho_\textnormal{de}$}

The system reduces to
\begin{align}
(1+z)\frac{d\rho_\text{c}}{dz} - 3\rho_\text{c} &= \beta \rho_\text{de}, \\
\frac{d\rho_\text{de}}{dz} &= \left[\frac{3(1+w_0)-\beta}{1+z} + \frac{3w_a z}{(1+z)^2}\right]\rho_\text{de}.
\end{align}

\subsubsection{Solution for $\rho_\textnormal{de}$}

The equation for $\rho_\text{de}$ can be written as
\begin{equation}
\frac{1}{\rho_\text{de}}\frac{d\rho_\text{de}}{dz}
= \frac{3(1+w_0)-\beta}{1+z} + \frac{3w_a z}{(1+z)^2}.
\end{equation}
Integrating,
\begin{equation}
\ln \rho_\text{de} = (3(1+w_0)-\beta)\ln(1+z)
+ \int \frac{3w_a z}{(1+z)^2}dz.
\end{equation}
Using the change of variable $u=1+z$,
\begin{align}
\int \frac{3w_a z}{(1+z)^2}dz
&= 3w_a \int \left(\frac{1}{u} - \frac{1}{u^2}\right)du \\
&= 3w_a \left[\ln(1+z) + \frac{1}{1+z}\right].
\end{align}
Thus,
\begin{equation}
\rho_\text{de}(z) = C (1+z)^{3(1+w_0+w_a)-\beta}
\exp\left(\frac{3w_a}{1+z}\right).
\end{equation}

Imposing $\rho_\text{de}(0)=\rho_\text{de,0}$ yields
\begin{equation}
C = \rho_\text{de,0} e^{-3w_a},
\end{equation}
and therefore
\begin{equation}
\rho_\text{de}(z) = \rho_\text{de,0}(1+z)^{3(1+w_0+w_a)-\beta}
\exp\left(-\frac{3w_a z}{1+z}\right).
\end{equation}

\subsubsection{Solution for $\rho_\textnormal{c}$}

Substituting into the dark matter equation,
\begin{multline}
\frac{d\rho_\text{c}}{dz} - \frac{3}{1+z}\rho_\text{c} \\
=
\beta \rho_\text{de,0}(1+z)^{3(w_0+w_a)-\beta+2}
\exp\left(-\frac{3w_a z}{1+z}\right).
\end{multline}

Using the integrating factor $\mu=(1+z)^{-3}$,
\begin{multline}
\frac{d}{dz}\left[(1+z)^{-3}\rho_\text{c}\right] \\
=
\beta \rho_\text{de,0}(1+z)^{3(w_0+w_a)-\beta-1}
\exp\left(-\frac{3w_a z}{1+z}\right).
\end{multline}
After the change of variable $x=1+z$, the integral on the right-hand side takes the form
\begin{equation}
I = e^{-3w_a} \int x^{K-1} \exp\left(\frac{3w_a}{x}\right) dx,
\end{equation}
with $K=3(w_0+w_a)-\beta$.

Using $u=-3w_a/x$, the integral can be expressed in terms of the incomplete gamma function, as
\begin{multline}
I = e^{-3w_a} (3w_a)(-3w_a)^{K-1} \\
\times 
\left[\Gamma(-K,-3w_a)
- \Gamma\left(-K,-\frac{3w_a}{1+z}\right)\right].
\end{multline}
This leads to the final solution for $\rho_\text{c}(z)$ given in Section \ref{subsec:CI}.

\subsection{Model CII: $Q = \beta H \rho_\textnormal{c}$}

The system becomes
\begin{align}
(1+z)\frac{d\rho_\text{c}}{dz} - 3\rho_\text{c} &= \beta \rho_\text{c}, \\
\frac{d\rho_\text{de}}{dz} &= \left[\frac{3(1+w_0)}{1+z} + \frac{3w_a z}{(1+z)^2}\right]\rho_\text{de}
- \frac{\beta \rho_\text{c}}{1+z}.
\end{align}

\subsubsection{Solution for $\rho_\textnormal{c}$}

This equation integrates directly,
\begin{equation}
\rho_\text{c}(z) = \rho_\text{c,0}(1+z)^{3+\beta}.
\end{equation}

\subsubsection{Solution for $\rho_\mathrm{de}$}

Substituting into the dark energy equation leads to
\begin{equation}
\frac{d \rho_\text{de}}{dz} - \left( \frac{3(1+w_{0})}{1+z} + \frac{3w_{a} z}{(1+z)^2}\right)\rho_\text{de} = - \beta  \rho_{c,0} (1+z)^{2+\beta}.
\end{equation}
This first-order linear differential equation can be solved by introducing the integrating factor
\begin{equation}
\mu(z) = (1+z)^{-3(1+w_0+w_a)}
\exp\left(-\frac{3w_a}{1+z}\right),
\end{equation}
which yields
\begin{multline}
\frac{d}{dz}\left[\mu(z)\rho_\text{de}\right]\\
=- \beta  \rho_\text{c,0} (1+z)^{- 3(w_{0} + w_{a})+\beta-1} \exp \left( \frac{-3w_a}{1+z}\right) .
\end{multline}

The resulting integral on the right-hand side can again be expressed in terms of incomplete gamma functions, providing the analytic solution presented in Section \ref{subsec:CII}.

\bibliography{sample}

@article{SupernovaSearchTeam:1998fmf,
    author = "Riess, Adam G. and others",
    collaboration = "Supernova Search Team",
    title = "{Observational evidence from supernovae for an accelerating universe and a cosmological constant}",
    eprint = "astro-ph/9805201",
    archivePrefix = "arXiv",
    doi = "10.1086/300499",
    journal = "Astron. J.",
    volume = "116",
    pages = "1009--1038",
    year = "1998"
}

@article{cosmicchronometers,
   title={Cosmic chronometers to calibrate the ladders and measure the curvature of the Universe. A model-independent study},
   volume={523},
   ISSN={1365-2966},
   url={http://dx.doi.org/10.1093/mnras/stad1621},
   DOI={10.1093/mnras/stad1621},
   number={3},
   journal={Monthly Notices of the Royal Astronomical Society},
   publisher={Oxford University Press (OUP)},
   author={Favale, Arianna and Gómez-Valent, Adrià and Migliaccio, Marina},
   year={2023},
   month=jun, pages={3406–3422} }

@misc{HZnewpoint1,
      title={Addressing the Hubble tension with cosmic chronometers}, 
      author={Michele Moresco},
      year={2023},
      eprint={2307.09501},
      archivePrefix={arXiv},
      primaryClass={astro-ph.CO},
      url={https://arxiv.org/abs/2307.09501}, 
}

@article{hznewpoint2,
   title={A new measurement of the expansion history of the Universe at z = 1.26 with cosmic chronometers in VANDELS},
   volume={679},
   ISSN={1432-0746},
   url={http://dx.doi.org/10.1051/0004-6361/202346992},
   DOI={10.1051/0004-6361/202346992},
   journal={Astronomy \& Astrophysics},
   publisher={EDP Sciences},
   author={Tomasetti, E. and Moresco, M. and Borghi, N. and Jiao, K. and Cimatti, A. and Pozzetti, L. and Carnall, A. C. and McLure, R. J. and Pentericci, L.},
   year={2023},
   month=nov, pages={A96} 
}

@article{hznewpoint3,
   title={New Observational H(z) Data from Full-spectrum Fitting of Cosmic Chronometers in the LEGA-C Survey},
   volume={265},
   ISSN={1538-4365},
   url={http://dx.doi.org/10.3847/1538-4365/acbc77},
   DOI={10.3847/1538-4365/acbc77},
   number={2},
   journal={The Astrophysical Journal Supplement Series},
   publisher={American Astronomical Society},
   author={Jiao, Kang and Borghi, Nicola and Moresco, Michele and Zhang, Tong-Jie},
   year={2023},
   month=mar, pages={48} }

@article{DATASN,
    author = "Scolnic, Dan and others",
    title = "{The Pantheon+ Analysis: The Full Data Set and Light-curve Release}",
    eprint = "2112.03863",
    archivePrefix = "arXiv",
    primaryClass = "astro-ph.CO",
    doi = "10.3847/1538-4357/ac8b7a",
    journal = "Astrophys. J.",
    volume = "938",
    number = "2",
    pages = "113",
    year = "2022"
}

@article{SupernovaCosmologyProject:1998vns,
    author = "Perlmutter, S. and others",
    collaboration = "Supernova Cosmology Project",
    title = "{Measurements of $\Omega$ and $\Lambda$ from 42 High Redshift Supernovae}",
    eprint = "astro-ph/9812133",
    archivePrefix = "arXiv",
    reportNumber = "LBNL-41801, LBL-41801",
    doi = "10.1086/307221",
    journal = "Astrophys. J.",
    volume = "517",
    pages = "565--586",
    year = "1999"
}

@article{Hu_1996,
   title={Small‐Scale Cosmological Perturbations: An Analytic Approach},
   volume={471},
   ISSN={1538-4357},
   url={http://dx.doi.org/10.1086/177989},
   DOI={10.1086/177989},
   number={2},
   journal={The Astrophysical Journal},
   publisher={American Astronomical Society},
   author={Hu, Wayne and Sugiyama, Naoshi},
   year={1996},
   month=nov, pages={542–570} 
}

@article{BAOFORMULA,
    author = {Eisenstein,Daniel J. and Hu, Wayne},
    title = {Baryonic Features in the Matter Transfer Function},
    journal = {The Astrophysical Journal},
    year = {1998},
    volume = {496},
    pages = {L2},
    publisher = {IOP Puslishing},
    adsurl = {https://iopscience.iop.org/article/10.1086/305424}
}

@article{DATABAO,
    author = {C. Nunes,Rafael and K. Yada, Santosh  and Jesus, J.F. and Bernui,Armando},
    title = {Cosmological parameter analyses using transversal BAO data},
    journal = {Monthly Notices of the Royal Astronomical Society},
    volume= {497},
    pages = {2133–2141},
    year =  {2020},
    adsurl = {https://academic.oup.com/mnras/article/497/2/2133/5870123},
}

@article{Bernal2020,
  author  = {Bernal, Jose Luis and Smith, Tristan L. and Boddy, Kimberly K. and Kamionkowski, Marc},
  title   = {Robustness of baryon acoustic oscillation constraints for early-Universe modifications to $\Lambda$CDM},
  journal = {Physical Review D},
  year    = {2020},
  volume  = {102},
  number  = {12},
  pages   = {123515},
  doi     = {10.1103/PhysRevD.102.123515}
}

@misc{desi2025,
    author = "Abdul Karim, M. and others",
    collaboration = "DESI",
    title = "{DESI DR2 Results II: Measurements of Baryon Acoustic Oscillations and Cosmological Constraints}",
    eprint = "2503.14738",
    archivePrefix = "arXiv",
    primaryClass = "astro-ph.CO",
    reportNumber = "FERMILAB-PUB-25-0169-PPD",
    month = "3",
    year = "2025"
}

@article{Pan-STARRS1:2017jku,
    author = "Scolnic, D. M. and others",
    collaboration = "Pan-STARRS1",
    title = "{The Complete Light-curve Sample of Spectroscopically Confirmed SNe Ia from Pan-STARRS1 and Cosmological Constraints from the Combined Pantheon Sample}",
    eprint = "1710.00845",
    archivePrefix = "arXiv",
    primaryClass = "astro-ph.CO",
    doi = "10.3847/1538-4357/aab9bb",
    journal = "Astrophys. J.",
    volume = "859",
    number = "2",
    pages = "101",
    year = "2018"
}

@article{Planck:2018vyg,
    author = "Aghanim, N. and others",
    collaboration = "Planck",
    title = "{Planck 2018 results. VI. Cosmological parameters}",
    eprint = "1807.06209",
    archivePrefix = "arXiv",
    primaryClass = "astro-ph.CO",
    doi = "10.1051/0004-6361/201833910",
    journal = "Astron. Astrophys.",
    volume = "641",
    pages = "A6",
    year = "2020",
    note = "[Erratum: \href{ https://doi.org/10.1051/0004-6361/201833910e}{Astron. Astrophys. \textbf{652}, C4(E) (2021)}]"
}

@article{SDSS:2005xqv,
    author = "Eisenstein, Daniel J. and others",
    collaboration = "SDSS",
    title = "{Detection of the Baryon Acoustic Peak in the Large-Scale Correlation Function of SDSS Luminous Red Galaxies}",
    eprint = "astro-ph/0501171",
    archivePrefix = "arXiv",
    reportNumber = "FERMILAB-PUB-05-057-A-CD",
    doi = "10.1086/466512",
    journal = "Astrophys. J.",
    volume = "633",
    pages = "560--574",
    year = "2005"
}

@article{eBOSS:2020yzd,
    author = "Alam, Shadab and others",
    collaboration = "eBOSS",
    title = "{Completed SDSS-IV extended Baryon Oscillation Spectroscopic Survey: Cosmological implications from two decades of spectroscopic surveys at the Apache Point Observatory}",
    eprint = "2007.08991",
    archivePrefix = "arXiv",
    primaryClass = "astro-ph.CO",
    doi = "10.1103/PhysRevD.103.083533",
    journal = "Phys. Rev. D",
    volume = "103",
    number = "8",
    pages = "083533",
    year = "2021"
}

@article{DESI:2024mwx,
    author = "Adame, A. G. and others",
    collaboration = "DESI",
    title = "{DESI 2024 VI: cosmological constraints from the measurements of baryon acoustic oscillations}",
    eprint = "2404.03002",
    archivePrefix = "arXiv",
    primaryClass = "astro-ph.CO",
    reportNumber = "FERMILAB-PUB-24-0154-PPD",
    doi = "10.1088/1475-7516/2025/02/021",
    journal = "JCAP",
    volume = "02",
    pages = "021",
    year = "2025"
}

@article{Copeland:2006wr,
    author = "Copeland, Edmund J. and Sami, M. and Tsujikawa, Shinji",
    title = "{Dynamics of dark energy}",
    eprint = "hep-th/0603057",
    archivePrefix = "arXiv",
    doi = "10.1142/S021827180600942X",
    journal = "Int. J. Mod. Phys. D",
    volume = "15",
    pages = "1753--1936",
    year = "2006"
}

@ARTICLE{CMB_DATA,
   author={{Chen}, Lu and {Huang}, Qing-Guo and {Wang}, Ke},
   title={Distance priors from Planck final release},
   journal={JCAP},
   year={2019},
   month=feb,
   volume={2019},
   number={2},
   eid={028},
   pages={028},
   doi={10.1088/1475-7516/2019/02/028},
   archivePrefix={arXiv},
   eprint={1808.05724},
   primaryClass={astro-ph.CO}
}

@article{SHIFT_PARAMETER,
   author={{Bond}, J.~R. and {Efstathiou}, G. and {Tegmark}, M.},
   title={Forecasting cosmic parameter errors from microwave background anisotropy experiments},
   journal={MNRAS},
   year={1997},
   month=nov,
   volume={291},
   number={3},
   pages={L33-L41},
   doi={10.1093/mnras/291.1.L33},
   archivePrefix={arXiv},
   eprint={astro-ph/9702100},
   primaryClass={astro-ph}
}

@article{Weinberg:1988cp,
    author = "Weinberg, Steven",
    editor = "Hsu, Jong-Ping and Fine, D.",
    title = "{The Cosmological Constant Problem}",
    reportNumber = "UTTG-12-88",
    doi = "10.1103/RevModPhys.61.1",
    journal = "Rev. Mod. Phys.",
    volume = "61",
    pages = "1--23",
    year = "1989"
}

@article{Martin:2012bt,
    author = "Martin, Jerome",
    title = "{Everything You Always Wanted To Know About The Cosmological Constant Problem (But Were Afraid To Ask)}",
    eprint = "1205.3365",
    archivePrefix = "arXiv",
    primaryClass = "astro-ph.CO",
    doi = "10.1016/j.crhy.2012.04.008",
    journal = "Comptes Rendus Physique",
    volume = "13",
    pages = "566--665",
    year = "2012"
}

@article{Riess:2019cxk,
    author = "Riess, Adam G. and Casertano, Stefano and Yuan, Wenlong and Macri, Lucas M. and Scolnic, Dan",
    title = "{Large Magellanic Cloud Cepheid Standards Provide a 1{\%} Foundation for the Determination of the Hubble Constant and Stronger Evidence for Physics beyond $\Lambda$CDM}",
    eprint = "1903.07603",
    archivePrefix = "arXiv",
    primaryClass = "astro-ph.CO",
    doi = "10.3847/1538-4357/ab1422",
    journal = "Astrophys. J.",
    volume = "876",
    number = "1",
    pages = "85",
    year = "2019"
}

@article{Riess:2021jrx,
    author = "Riess, Adam G. and others",
    title = "{A Comprehensive Measurement of the Local Value of the Hubble Constant with 1 km s$^{-1}$ Mpc$^{-1}$ Uncertainty from the Hubble Space Telescope and the SH0ES Team}",
    eprint = "2112.04510",
    archivePrefix = "arXiv",
    primaryClass = "astro-ph.CO",
    doi = "10.3847/2041-8213/ac5c5b",
    journal = "Astrophys. J. Lett.",
    volume = "934",
    number = "1",
    pages = "L7",
    year = "2022"
}

@article{DES:2017myr,
    author = "Abbott, T. M. C. and others",
    collaboration = "DES",
    title = "{Dark Energy Survey year 1 results: Cosmological constraints from galaxy clustering and weak lensing}",
    eprint = "1708.01530",
    archivePrefix = "arXiv",
    primaryClass = "astro-ph.CO",
    reportNumber = "FERMILAB-PUB-17-294-PPD",
    doi = "10.1103/PhysRevD.98.043526",
    journal = "Phys. Rev. D",
    volume = "98",
    number = "4",
    pages = "043526",
    year = "2018"
}

@article{Heymans:2020gsg,
    author = "Heymans, Catherine and others",
    title = "{KiDS-1000 Cosmology: Multi-probe weak gravitational lensing and spectroscopic galaxy clustering constraints}",
    eprint = "2007.15632",
    archivePrefix = "arXiv",
    primaryClass = "astro-ph.CO",
    doi = "10.1051/0004-6361/202039063",
    journal = "Astron. Astrophys.",
    volume = "646",
    pages = "A140",
    year = "2021"
}

@article{Verde:2019ivm,
    author = "Verde, L. and Treu, T. and Riess, A. G.",
    title = "{Tensions between the Early and the Late Universe}",
    eprint = "1907.10625",
    archivePrefix = "arXiv",
    primaryClass = "astro-ph.CO",
    doi = "10.1038/s41550-019-0902-0",
    journal = "Nature Astron.",
    volume = "3",
    pages = "891",
    year = "2019"
}

@article{DiValentino:2020zio,
    author = "Di Valentino, Eleonora and others",
    title = "{Snowmass2021 - Letter of interest cosmology intertwined II: The hubble constant tension}",
    eprint = "2008.11284",
    archivePrefix = "arXiv",
    primaryClass = "astro-ph.CO",
    reportNumber = "FERMILAB-PUB-21-590-PPD",
    doi = "10.1016/j.astropartphys.2021.102605",
    journal = "Astropart. Phys.",
    volume = "131",
    pages = "102605",
    year = "2021"
}

@article{DiValentino:2020vvd,
    author = "Di Valentino, Eleonora and others",
    title = "{Cosmology Intertwined III: $f \sigma_8$ and $S_8$}",
    eprint = "2008.11285",
    archivePrefix = "arXiv",
    primaryClass = "astro-ph.CO",
    reportNumber = "FERMILAB-PUB-20-495-AE",
    doi = "10.1016/j.astropartphys.2021.102604",
    journal = "Astropart. Phys.",
    volume = "131",
    pages = "102604",
    year = "2021"
}

@article{DiValentino:2021izs,
    author = "Di Valentino, Eleonora and Mena, Olga and Pan, Supriya and Visinelli, Luca and Yang, Weiqiang and Melchiorri, Alessandro and Mota, David F. and Riess, Adam G. and Silk, Joseph",
    title = "{In the realm of the Hubble tension{\textemdash}a review of solutions}",
    eprint = "2103.01183",
    archivePrefix = "arXiv",
    primaryClass = "astro-ph.CO",
    reportNumber = "IPPP/20/108",
    doi = "10.1088/1361-6382/ac086d",
    journal = "Class. Quant. Grav.",
    volume = "38",
    number = "15",
    pages = "153001",
    year = "2021"
}

@article{Perivolaropoulos:2021jda,
    author = "Perivolaropoulos, Leandros and Skara, Foteini",
    title = "{Challenges for {\ensuremath{\Lambda}}CDM: An update}",
    eprint = "2105.05208",
    archivePrefix = "arXiv",
    primaryClass = "astro-ph.CO",
    doi = "10.1016/j.newar.2022.101659",
    journal = "New Astron. Rev.",
    volume = "95",
    pages = "101659",
    year = "2022"
}

@article{Nojiri:2006ri,
    author = "Nojiri, Shin'ichi and Odintsov, Sergei D.",
    editor = "Borowiec, Andrzej",
    title = "{Introduction to modified gravity and gravitational alternative for dark energy}",
    eprint = "hep-th/0601213",
    archivePrefix = "arXiv",
    reportNumber = "KARP-2006-06",
    doi = "10.1142/S0219887807001928",
    journal = "eConf",
    volume = "C0602061",
    pages = "06",
    year = "2006"
}

@article{Clifton:2011jh,
    author = "Clifton, Timothy and Ferreira, Pedro G. and Padilla, Antonio and Skordis, Constantinos",
    title = "{Modified Gravity and Cosmology}",
    eprint = "1106.2476",
    archivePrefix = "arXiv",
    primaryClass = "astro-ph.CO",
    doi = "10.1016/j.physrep.2012.01.001",
    journal = "Phys. Rept.",
    volume = "513",
    pages = "1--189",
    year = "2012"
}

@article{Bamba2012,
    author = "Bamba, Kazuharu and Capozziello, Salvatore and Nojiri, Shin'ichi and Odintsov, Sergei D.",
    title = "{Dark energy cosmology: the equivalent description via different theoretical models and cosmography tests}",
    eprint = "1205.3421",
    archivePrefix = "arXiv",
    primaryClass = "gr-qc",
    doi = "10.1007/s10509-012-1181-8",
    journal = "Astrophys. Space Sci.",
    volume = "342",
    pages = "155--228",
    year = "2012"
}

@article{Joyce:2016vqv,
    author = "Joyce, Austin and Lombriser, Lucas and Schmidt, Fabian",
    title = "{Dark Energy Versus Modified Gravity}",
    eprint = "1601.06133",
    archivePrefix = "arXiv",
    primaryClass = "astro-ph.CO",
    doi = "10.1146/annurev-nucl-102115-044553",
    journal = "Ann. Rev. Nucl. Part. Sci.",
    volume = "66",
    pages = "95--122",
    year = "2016"
}

@article{Chevallier:2000qy,
    author = "Chevallier, Michel and Polarski, David",
    title = "{Accelerating universes with scaling dark matter}",
    eprint = "gr-qc/0009008",
    archivePrefix = "arXiv",
    doi = "10.1142/S0218271801000822",
    journal = "Int. J. Mod. Phys. D",
    volume = "10",
    pages = "213--224",
    year = "2001"
}

@article{Linder:2002et,
    author = "Linder, Eric V.",
    title = "{Exploring the expansion history of the universe}",
    eprint = "astro-ph/0208512",
    archivePrefix = "arXiv",
    doi = "10.1103/PhysRevLett.90.091301",
    journal = "Phys. Rev. Lett.",
    volume = "90",
    pages = "091301",
    year = "2003"
}

@article{DESI:2025zgx,
    author = "Abdul Karim, M. and others",
    collaboration = "DESI",
    title = "{DESI DR2 results. II. Measurements of baryon acoustic oscillations and cosmological constraints}",
    eprint = "2503.14738",
    archivePrefix = "arXiv",
    primaryClass = "astro-ph.CO",
    reportNumber = "FERMILAB-PUB-25-0169-PPD",
    doi = "10.1103/tr6y-kpc6",
    journal = "Phys. Rev. D",
    volume = "112",
    number = "8",
    pages = "083515",
    year = "2025"
}

@article{Du:2024pai,
    author = "Du, Guo-Hong and Wu, Peng-Ju and Li, Tian-Nuo and Zhang, Xin",
    title = "{Impacts of dark energy on weighing neutrinos after DESI BAO}",
    eprint = "2407.15640",
    archivePrefix = "arXiv",
    primaryClass = "astro-ph.CO",
    doi = "10.1140/epjc/s10052-025-14094-0",
    journal = "Eur. Phys. J. C",
    volume = "85",
    number = "4",
    pages = "392",
    year = "2025"
}

@article{Giare:2024oil,
    author = "Giar{\`e}, William",
    title = "{Dynamical dark energy beyond Planck? Constraints from multiple CMB probes, DESI BAO, and type-Ia supernovae}",
    eprint = "2409.17074",
    archivePrefix = "arXiv",
    primaryClass = "astro-ph.CO",
    doi = "10.1103/ss37-cxhn",
    journal = "Phys. Rev. D",
    volume = "112",
    number = "2",
    pages = "023508",
    year = "2025"
}

@article{Zheng:2024qzi,
    author = "Zheng, Jie and Qiang, Da-Chun and You, Zhi-Qiang",
    title = "{Cosmological constraints on dark energy models using DESI BAO 2024}",
    eprint = "2412.04830",
    archivePrefix = "arXiv",
    primaryClass = "astro-ph.CO",
    doi = "10.1088/1475-7516/2025/08/056",
    journal = "JCAP",
    volume = "08",
    pages = "056",
    year = "2025"
}

@article{Ormondroyd:2025iaf,
    author = {Ormondroyd, A N and Handley, W J and Hobson, M P and Lasenby, A N},
    title = {Comparison of dynamical dark energy with {$\Lambda$CDM} in light of DESI DR2},
    journal = {Monthly Notices of the Royal Astronomical Society},
    volume = {547},
    number = {2},
    pages = {staf2207},
    year = {2026},
    month = {04},
    issn = {0035-8711},
    doi = {10.1093/mnras/staf2207},
    url = {https://doi.org/10.1093/mnras/staf2207},
       eprint = "2503.17342",
    archivePrefix = "arXiv"
}

@article{Nesseris:2025lke,
  title = {What we have not learned about the dark energy equation of state},
  author = {Nesseris, Savvas and Akrami, Yashar and Starkman, Glenn D.},
  journal = {Phys. Rev. D},
  volume = {114},
  issue = {4},
  pages = {L041305},
  numpages = {6},
  year = {2026},
  month = {Aug},
  publisher = {American Physical Society},
  doi = {10.1103/q2gb-n61g},
  url = {https://link.aps.org/doi/10.1103/q2gb-n61g},
  eprint = "2503.22529",
    archivePrefix = "arXiv",
    primaryClass = "astro-ph.CO",
}

@article{DESI:2025wyn,
    author = "Gu, Gan and others",
    collaboration = "DESI",
    title = "{Dynamical dark energy in light of the DESI DR2 baryonic acoustic oscillations measurements}",
    eprint = "2504.06118",
    archivePrefix = "arXiv",
    primaryClass = "astro-ph.CO",
    reportNumber = "FERMILAB-PUB-25-0235-PPD",
    doi = "10.1038/s41550-025-02669-6",
    journal = "Nature Astron.",
    volume = "9",
    number = "12",
    pages = "1879--1889",
    year = "2025",
    note = "[Erratum: Nature Astron. 9, 1898 (2025)]"
}

@article{Scherer:2025esj,
    author = "Scherer, Mateus and Sabogal, Miguel A. and Nunes, Rafael C. and De Felice, Antonio",
    title = "{Challenging the {\ensuremath{\Lambda}}CDM model: 5{\ensuremath{\sigma}} evidence for a dynamical dark energy late-time transition}",
    eprint = "2504.20664",
    archivePrefix = "arXiv",
    primaryClass = "astro-ph.CO",
    doi = "10.1103/n86r-sjgm",
    journal = "Phys. Rev. D",
    volume = "112",
    number = "4",
    pages = "043513",
    year = "2025"
}

@article{Capozziello:2025qmh,
    author = "Capozziello, Salvatore and Chaudhary, Himanshu and Harko, Tiberiu and Mustafa, Ghulam",
    title = "{Is dark energy dynamical in the DESI era? A critical review}",
    eprint = "2512.10585",
    archivePrefix = "arXiv",
    primaryClass = "astro-ph.CO",
    doi = "10.1016/j.dark.2025.102196",
    journal = "Phys. Dark Univ.",
    volume = "51",
    pages = "102196",
    year = "2026"
}

@article{Wetterich:1994bg,
    author = "Wetterich, Christof",
    title = "{The Cosmon model for an asymptotically vanishing time dependent cosmological 'constant'}",
    eprint = "hep-th/9408025",
    archivePrefix = "arXiv",
    reportNumber = "HD-THEP-94-16",
    journal = "Astron. Astrophys.",
    volume = "301",
    pages = "321--328",
    year = "1995"
}

@article{Amendola:1999er,
    author = "Amendola, Luca",
    title = "{Coupled quintessence}",
    eprint = "astro-ph/9908023",
    archivePrefix = "arXiv",
    doi = "10.1103/PhysRevD.62.043511",
    journal = "Phys. Rev. D",
    volume = "62",
    pages = "043511",
    year = "2000"
}

@article{Zimdahl:2001ar,
    author = "Zimdahl, Winfried and Pavon, Diego",
    title = "{Interacting quintessence}",
    eprint = "astro-ph/0105479",
    archivePrefix = "arXiv",
    doi = "10.1016/S0370-2693(01)01174-1",
    journal = "Phys. Lett. B",
    volume = "521",
    pages = "133--138",
    year = "2001"
}

@article{Chimento:2003iea,
    author = "Chimento, Luis P. and Jakubi, Alejandro S. and Pavon, Diego and Zimdahl, Winfried",
    title = "{Interacting quintessence solution to the coincidence problem}",
    eprint = "astro-ph/0303145",
    archivePrefix = "arXiv",
    doi = "10.1103/PhysRevD.67.083513",
    journal = "Phys. Rev. D",
    volume = "67",
    pages = "083513",
    year = "2003"
}

@article{Farrar:2003uw,
    author = "Farrar, Glennys R. and Peebles, P. James E.",
    title = "{Interacting dark matter and dark energy}",
    eprint = "astro-ph/0307316",
    archivePrefix = "arXiv",
    doi = "10.1086/381728",
    journal = "Astrophys. J.",
    volume = "604",
    pages = "1--11",
    year = "2004"
}

@article{Valiviita:2008iv,
    author = "Valiviita, Jussi and Majerotto, Elisabetta and Maartens, Roy",
    title = "{Instability in interacting dark energy and dark matter fluids}",
    eprint = "0804.0232",
    archivePrefix = "arXiv",
    primaryClass = "astro-ph",
    doi = "10.1088/1475-7516/2008/07/020",
    journal = "JCAP",
    volume = "07",
    pages = "020",
    year = "2008"
}

@article{He:2008si,
    author = "He, Jian-Hua and Wang, Bin and Abdalla, Elcio",
    title = "{Stability of the curvature perturbation in dark sectors' mutual interacting models}",
    eprint = "0807.3471",
    archivePrefix = "arXiv",
    primaryClass = "gr-qc",
    doi = "10.1016/j.physletb.2008.11.062",
    journal = "Phys. Lett. B",
    volume = "671",
    pages = "139--145",
    year = "2009"
}

@article{HeWang2008,
  author  = {He, Jian-Hua and Wang, Bin},
  title   = {Effects of the Interaction between Dark Energy and Dark Matter on Cosmological Parameters},
  journal = {JCAP},
  volume  = {06},
  pages   = {010},
  year    = {2008},
  eprint  = {0801.4233},
  archivePrefix = {arXiv}
}

@article{Jackson:2009mz,
    author = "Jackson, Brendan M. and Taylor, Andy and Berera, Arjun",
    title = "{On the large-scale instability in interacting dark energy and dark matter fluids}",
    eprint = "0901.3272",
    archivePrefix = "arXiv",
    primaryClass = "astro-ph.CO",
    doi = "10.1103/PhysRevD.79.043526",
    journal = "Phys. Rev. D",
    volume = "79",
    pages = "043526",
    year = "2009"
}

@article{Wang2016,
  author  = {Wang, Bin and Abdalla, Elcio and Atrio-Barandela, Fernando and Pav{\'o}n, Diego},
  title   = {Dark Matter and Dark Energy Interactions: Theoretical Challenges, Cosmological Implications and Observational Signatures},
  journal = {Reports on Progress in Physics},
  volume  = {79},
  number  = {9},
  pages   = {096901},
  year    = {2016},
  doi     = {10.1088/0034-4885/79/9/096901},
  eprint  = {1603.08299},
  archivePrefix = {arXiv},
  primaryClass  = {astro-ph.CO}
}

@article{Bolotin2015,
  author  = {Bolotin, Yuri L. and Kostenko, Alexander and Lemets, Oleg A. and Yerokhin, Danylo A.},
  title   = {Cosmological Evolution With Interaction Between Dark Energy And Dark Matter},
  journal = {International Journal of Modern Physics D},
  volume  = {24},
  number  = {03},
  pages   = {1530007},
  year    = {2015},
  doi     = {10.1142/S0218271815300074},
  eprint  = {1310.0085},
  archivePrefix = {arXiv},
  primaryClass  = {gr-qc}
}

@article{DiValentino:2017iww,
    author = "Di Valentino, Eleonora and Melchiorri, Alessandro and Mena, Olga",
    title = "{Can interacting dark energy solve the $H_0$ tension?}",
    eprint = "1704.08342",
    archivePrefix = "arXiv",
    primaryClass = "astro-ph.CO",
    doi = "10.1103/PhysRevD.96.043503",
    journal = "Phys. Rev. D",
    volume = "96",
    number = "4",
    pages = "043503",
    year = "2017"
}

@article{Yang:2018uae,
    author = "Yang, Weiqiang and Mukherjee, Ankan and Di Valentino, Eleonora and Pan, Supriya",
    title = "{Interacting dark energy with time varying equation of state and the $H_0$ tension}",
    eprint = "1809.06883",
    archivePrefix = "arXiv",
    primaryClass = "astro-ph.CO",
    doi = "10.1103/PhysRevD.98.123527",
    journal = "Phys. Rev. D",
    volume = "98",
    number = "12",
    pages = "123527",
    year = "2018"
}

@article{Yang:2018euj,
    author = "Yang, Weiqiang and Pan, Supriya and Di Valentino, Eleonora and Nunes, Rafael C. and Vagnozzi, Sunny and Mota, David F.",
    title = "{Tale of stable interacting dark energy, observational signatures, and the $H_0$ tension}",
    eprint = "1805.08252",
    archivePrefix = "arXiv",
    primaryClass = "astro-ph.CO",
    doi = "10.1088/1475-7516/2018/09/019",
    journal = "JCAP",
    volume = "09",
    pages = "019",
    year = "2018"
}

@article{Pan:2019gop,
    author = "Pan, Supriya and Yang, Weiqiang and Di Valentino, Eleonora and Saridakis, Emmanuel N. and Chakraborty, Subenoy",
    title = "{Interacting scenarios with dynamical dark energy: Observational constraints and alleviation of the $H_0$ tension}",
    eprint = "1907.07540",
    archivePrefix = "arXiv",
    primaryClass = "astro-ph.CO",
    doi = "10.1103/PhysRevD.100.103520",
    journal = "Phys. Rev. D",
    volume = "100",
    number = "10",
    pages = "103520",
    year = "2019"
}

@article{DiValentino:2019ffd,
    author = "Di Valentino, Eleonora and Melchiorri, Alessandro and Mena, Olga and Vagnozzi, Sunny",
    title = "{Interacting dark energy in the early 2020s: A promising solution to the $H_0$ and cosmic shear tensions}",
    eprint = "1908.04281",
    archivePrefix = "arXiv",
    primaryClass = "astro-ph.CO",
    doi = "10.1016/j.dark.2020.100666",
    journal = "Phys. Dark Univ.",
    volume = "30",
    pages = "100666",
    year = "2020"
}

@article{Wang:2024vmw,
    author = "Wang, B. and Abdalla, E. and Atrio-Barandela, F. and Pav{\'o}n, D.",
    title = "{Further understanding the interaction between dark energy and dark matter: current status and future directions}",
    eprint = "2402.00819",
    archivePrefix = "arXiv",
    primaryClass = "astro-ph.CO",
    doi = "10.1088/1361-6633/ad2527",
    journal = "Rept. Prog. Phys.",
    volume = "87",
    number = "3",
    pages = "036901",
    year = "2024"
}

@article{Gavela:2009cy,
    author = "Gavela, M. B. and Hernandez, D. and Lopez Honorez, L. and Mena, O. and Rigolin, S.",
    title = "{Dark coupling}",
    eprint = "0901.1611",
    archivePrefix = "arXiv",
    primaryClass = "astro-ph.CO",
    reportNumber = "FTUAM-08-26, FT-UAM-CSIC-08-94, DFPD-09-TH-01, ULB-TH-09-01, IFIC-09-01",
    doi = "10.1088/1475-7516/2009/07/034",
    journal = "JCAP",
    volume = "07",
    pages = "034",
    year = "2009",
    note = "[Erratum: \href{ https://doi.org/10.1088/1475-7516/2010/05/E01}{JCAP \textbf{05}, E01 (2010)}]"
}

@article{Costa:2013sva,
    author = "Costa, Andr{\'e} A. and Xu, Xiao-Dong and Wang, Bin and Ferreira, Elisa G. M. and Abdalla, E.",
    title = "{Testing the Interaction between Dark Energy and Dark Matter with Planck Data}",
    eprint = "1311.7380",
    archivePrefix = "arXiv",
    primaryClass = "astro-ph.CO",
    doi = "10.1103/PhysRevD.89.103531",
    journal = "Phys. Rev. D",
    volume = "89",
    number = "10",
    pages = "103531",
    year = "2014"
}

@article{Pan:2012ki,
    author = "Pan, Supriya and Bhattacharya, Subhra and Chakraborty, Subenoy",
    title = "{An analytic model for interacting dark energy and its observational constraints}",
    eprint = "1210.0396",
    archivePrefix = "arXiv",
    primaryClass = "gr-qc",
    doi = "10.1093/mnras/stv1495",
    journal = "Mon. Not. Roy. Astron. Soc.",
    volume = "452",
    number = "3",
    pages = "3038--3046",
    year = "2015"
}

@article{vanderWesthuizen:2025vcb,
    author = "van der Westhuizen, Marcel and Abebe, Amare and Di Valentino, Eleonora",
    title = "{I. Linear interacting dark energy: Analytical solutions and theoretical pathologies}",
    eprint = "2509.04495",
    archivePrefix = "arXiv",
    primaryClass = "gr-qc",
    doi = "10.1016/j.dark.2025.102119",
    journal = "Phys. Dark Univ.",
    volume = "50",
    pages = "102119",
    year = "2025"
}

@article{vanderWesthuizen:2025mnw,
    author = "van der Westhuizen, Marcel and Abebe, Amare and Di Valentino, Eleonora",
    title = "{II. Non-linear interacting dark energy: Analytical solutions and theoretical pathologies}",
    eprint = "2509.04494",
    archivePrefix = "arXiv",
    primaryClass = "gr-qc",
    doi = "10.1016/j.dark.2025.102120",
    journal = "Phys. Dark Univ.",
    volume = "50",
    pages = "102120",
    year = "2025"
}

@article{vanderWesthuizen:2025rip,
    author = "van der Westhuizen, Marcel and Abebe, Amare and Di Valentino, Eleonora",
    title = "{III. Interacting Dark Energy: Summary of models, Pathologies, and Constraints}",
    eprint = "2509.04496",
    archivePrefix = "arXiv",
    primaryClass = "gr-qc",
    doi = "10.1016/j.dark.2025.102121",
    journal = "Phys. Dark Univ.",
    volume = "50",
    pages = "102121",
    year = "2025"
}

@article{Clemson:2011an,
    author = "Clemson, Timothy and Koyama, Kazuya and Zhao, Gong-Bo and Maartens, Roy and Valiviita, Jussi",
    title = "{Interacting Dark Energy -- constraints and degeneracies}",
    eprint = "1109.6234",
    archivePrefix = "arXiv",
    primaryClass = "astro-ph.CO",
    doi = "10.1103/PhysRevD.85.043007",
    journal = "Phys. Rev. D",
    volume = "85",
    pages = "043007",
    year = "2012"
}

@article{Li:2014eha,
    author = "Li, Yun-He and Zhang, Jing-Fei and Zhang, Xin",
    title = "{Parametrized Post-Friedmann Framework for Interacting Dark Energy}",
    eprint = "1404.5220",
    archivePrefix = "arXiv",
    primaryClass = "astro-ph.CO",
    doi = "10.1103/PhysRevD.90.063005",
    journal = "Phys. Rev. D",
    volume = "90",
    number = "6",
    pages = "063005",
    year = "2014"
}

@article{vanderWesthuizen:2023hcl,
    author = "van der Westhuizen, Marcel A. and Abebe, Amare",
    title = "{Interacting dark energy: clarifying the cosmological implications and viability conditions}",
    eprint = "2302.11949",
    archivePrefix = "arXiv",
    primaryClass = "gr-qc",
    doi = "10.1088/1475-7516/2024/01/048",
    journal = "JCAP",
    volume = "01",
    pages = "048",
    year = "2024"
}

@article{Hoerning:2023hks,
    author = "Hoerning, Gabriel A. and Landim, Ricardo G. and Ponte, Luiza O. and Rolim, Raphael P. and Abdalla, Filipe B. and Abdalla, Elcio",
    title = "{Constraints on interacting dark energy revisited: Implications for the Hubble tension}",
    eprint = "2308.05807",
    archivePrefix = "arXiv",
    primaryClass = "astro-ph.CO",
    doi = "10.1103/6zrh-8fmv",
    journal = "Phys. Rev. D",
    volume = "112",
    number = "2",
    pages = "023523",
    year = "2025"
}

@article{Benisty:2024lmj,
    author = "Benisty, David and Pan, Supriya and Staicova, Denitsa and Di Valentino, Eleonora and Nunes, Rafael C.",
    title = "{Late-time constraints on interacting dark energy: Analysis independent of H0, rd, and MB}",
    eprint = "2403.00056",
    archivePrefix = "arXiv",
    primaryClass = "astro-ph.CO",
    doi = "10.1051/0004-6361/202449883",
    journal = "Astron. Astrophys.",
    volume = "688",
    pages = "A156",
    year = "2024"
}

@article{Giare:2024smz,
    author = "Giar{\`e}, William and Sabogal, Miguel A. and Nunes, Rafael C. and Di Valentino, Eleonora",
    title = "{Interacting Dark Energy after DESI Baryon Acoustic Oscillation Measurements}",
    eprint = "2404.15232",
    archivePrefix = "arXiv",
    primaryClass = "astro-ph.CO",
    doi = "10.1103/PhysRevLett.133.251003",
    journal = "Phys. Rev. Lett.",
    volume = "133",
    number = "25",
    pages = "251003",
    year = "2024"
}

@article{Ghedini:2024mdu,
    author = "Ghedini, Pietro and Hajjar, Rasmi and Mena, Olga",
    title = "{Redshift-space distortions corner interacting dark energy}",
    eprint = "2409.02700",
    archivePrefix = "arXiv",
    primaryClass = "astro-ph.CO",
    doi = "10.1016/j.dark.2024.101671",
    journal = "Phys. Dark Univ.",
    volume = "46",
    pages = "101671",
    year = "2024"
}

@article{Zhu:2025lrk,
    author = "Zhu, Ziyan and Jiang, Qingquan and Liu, Yu and Wu, Puxun and Liang, Nan",
    title = "{Cosmological constraints on the phenomenological interacting dark energy model with Fermi gamma-ray bursts and DESI DR2}",
    eprint = "2511.16032",
    archivePrefix = "arXiv",
    primaryClass = "astro-ph.CO",
    doi = "10.1016/j.jheap.2025.100534",
    journal = "JHEAp",
    volume = "51",
    pages = "100534",
    year = "2026"
}

@article{Tsedrik:2025jdv,
  author        = {Tsedrik, M. and Lee, S. and Markovic, K. and Carrilho, P. and Pourtsidou, A. and Moretti, C. and Bose, B. and Huff, E. and Robertson, A. and Taylor, P. L. and Zuntz, J.},
  title         = {Interacting dark energy constraints from the full-shape analyses of BOSS DR12 and DES Year 3 measurements},
  journal       = {Monthly Notices of the Royal Astronomical Society: Letters},
  volume        = {541},
  number        = {1},
  pages         = {L65--L70},
  year          = {2025},
  doi           = {10.1093/mnrasl/slaf055},
  eprint        = {2502.03390},
  archivePrefix = {arXiv},
  primaryClass  = {astro-ph.CO}
}

@article{Petri:2025swg,
    author = "Petri, Vitor and Marra, Valerio and von Marttens, Rodrigo",
    title = "{Dark degeneracy in DESI DR2 data: Interacting or evolving dark energy?}",
    eprint = "2508.17955",
    archivePrefix = "arXiv",
    primaryClass = "astro-ph.CO",
    doi = "10.1103/3k93-p1n8",
    journal = "Phys. Rev. D",
    volume = "113",
    number = "2",
    pages = "023504",
    year = "2026"
}

@article{Figueruelo:2026eis,
    author = "Figueruelo, David and van der Westhuizen, Marcel and Abebe, Amare and Di Valentino, Eleonora",
    title = "{Late-time background constraints on linear and non-linear interacting dark energy after DESI DR2}",
    eprint = "2601.03122",
    archivePrefix = "arXiv",
    primaryClass = "astro-ph.CO",
    doi = "10.1016/j.dark.2026.102238",
    journal = "Phys. Dark Univ.",
    volume = "52",
    pages = "102238",
    year = "2026"
}

@misc{Li:2026xaz,
    author = "Li, Tian-Nuo and Giar{\`e}, William and Du, Guo-Hong and Li, Yun-He and Di Valentino, Eleonora and Zhang, Jing-Fei and Zhang, Xin",
    title = "{Strong Evidence for Dark Sector Interactions}",
    eprint = "2601.07361",
    archivePrefix = "arXiv",
    primaryClass = "astro-ph.CO",
    month = "1",
    year = "2026"
}

@article{Xu:2011tsa,
    author = "Xu, Xiao-Dong and Wang, Bin",
    title = "{Breaking parameter degeneracy in interacting dark energy models from observations}",
    eprint = "1103.2632",
    archivePrefix = "arXiv",
    primaryClass = "astro-ph.CO",
    doi = "10.1016/j.physletb.2011.06.043",
    journal = "Phys. Lett. B",
    volume = "701",
    pages = "513--519",
    year = "2011"
}

@article{Carneiro:2014uua,
    author = "Carneiro, S. and Borges, H. A.",
    title = "{On dark degeneracy and interacting models}",
    eprint = "1402.2316",
    archivePrefix = "arXiv",
    primaryClass = "astro-ph.CO",
    doi = "10.1088/1475-7516/2014/06/010",
    journal = "JCAP",
    volume = "06",
    pages = "010",
    year = "2014"
}

@article{mcmchammer,
       author = {{Foreman-Mackey}, Daniel and {Hogg}, David W. and {Lang}, Dustin and {Goodman}, Jonathan},
        title = "{emcee: The MCMC Hammer}",
      journal = {pasp},
         year = 2013,
        month = mar,
       volume = {125},
       number = {925},
        pages = {306},
          doi = {10.1086/670067},
archivePrefix = {arXiv},
       eprint = {1202.3665},
 primaryClass = {astro-ph.IM},
       adsurl = {https://ui.adsabs.harvard.edu/abs/2013PASP..125..306F}
}

@ARTICLE{1974ITAC...19..716A,
       author = {{Akaike}, H.},
        title = "{A New Look at the Statistical Model Identification}",
      journal = {IEEE Transactions on Automatic Control},
         year = 1974,
        month = jan,
       volume = {19},
        pages = {716-723},
       adsurl = {https://ui.adsabs.harvard.edu/abs/1974ITAC...19..716A},
      doi = {10.1109/TAC.1974.1100705}
}

@incollection{Sokal1996MonteCM,
  author    = {Sokal, Alan D.},
  title     = {Monte Carlo Methods in Statistical Mechanics: Foundations and New Algorithms},
  booktitle = {Functional Integration},
  editor    = {DeWitt-Morette, C{\'e}cile and Cartier, Pierre and Folacci, Antoine},
  series    = {NATO ASI Series B: Physics},
  volume    = {361},
  pages     = {131--192},
  publisher = {Springer},
  address   = {Boston, MA},
  year      = {1997},
  doi       = {10.1007/978-1-4899-0319-8_6}
}

@article{BIC1974,
author = {Gideon Schwarz},
title = {{Estimating the Dimension of a Model}},
volume = {6},
journal = {The Annals of Statistics},
number = {2},
publisher = {Institute of Mathematical Statistics},
pages = {461 -- 464},
year = {1978},
doi = {10.1214/aos/1176344136},
URL = {https://doi.org/10.1214/aos/1176344136}
}

@article{Barrow2000,
    author = "Barrow, John and Bean, Rachel and Magueijo, Joao",
    title = "{Can the universe escape eternal acceleration?}",
    eprint = "astro-ph/0004321",
    archivePrefix = "arXiv",
    doi = "10.1046/j.1365-8711.2000.03778.x",
    journal = "Mon. Not. Roy. Astron. Soc.",
    volume = "316",
    pages = "L41",
    year = "2000"
}

@article{Vargas2011,
    author = "Vargas, Cristofher Zuniga and Hipolito-Ricaldi, Wiliam S. and Zimdahl, Winfried",
    title = "{Perturbations for transient acceleration}",
    eprint = "1112.5337",
    archivePrefix = "arXiv",
    primaryClass = "astro-ph.CO",
    doi = "10.1088/1475-7516/2012/04/032",
    journal = "JCAP",
    volume = "04",
    pages = "032",
    year = "2012"
}

@article{Shahalam2015,
    author = "Shahalam, M. and Sami, Sasha and Agarwal, Abhineet",
    title = "{$Om$ diagnostic applied to scalar field models and slowing down of cosmic acceleration}",
    eprint = "1501.04047",
    archivePrefix = "arXiv",
    primaryClass = "astro-ph.CO",
    doi = "10.1093/mnras/stv083",
    journal = "Mon. Not. Roy. Astron. Soc.",
    volume = "448",
    number = "3",
    pages = "2948--2959",
    year = "2015"
}

@article{Hu2015,
    author = "Hu, Yazhou and Li, Miao and Li, Nan and Wang, Shuang",
    title = "{A comprehensive investigation on the slowing down of cosmic acceleration}",
    eprint = "1509.03461",
    archivePrefix = "arXiv",
    primaryClass = "astro-ph.CO",
    doi = "10.3847/0004-637X/821/1/60",
    journal = "Astrophys. J.",
    volume = "821",
    number = "1",
    pages = "60",
    year = "2016"
}

@article{Magana2014,
    author = "Maga\~na, Juan and C\'ardenas, Victor H. and Motta, V.",
    title = "{Cosmic slowing down of acceleration for several dark energy parametrizations}",
    eprint = "1407.1632",
    archivePrefix = "arXiv",
    primaryClass = "astro-ph.CO",
    doi = "10.1088/1475-7516/2014/10/017",
    journal = "JCAP",
    volume = "10",
    pages = "017",
    year = "2014"
}

@article{Zhang2018,
    author = "Zhang, Ming-Jian and Xia, Jun-Qing",
    title = "{Physical condition for the slowing down of cosmic acceleration}",
    eprint = "1701.04973",
    archivePrefix = "arXiv",
    primaryClass = "astro-ph.CO",
    doi = "10.1016/j.nuclphysb.2018.02.020",
    journal = "Nucl. Phys. B",
    volume = "929",
    pages = "438--451",
    year = "2018"
}

@article{Bolotin2020,
    author = "Bolotin, Yu. L. and Cherkaskiy, V. A. and Konchatnyi, M. I. and Pan, Supriya and Yang, Weiqiang",
    title = "{Do current observations support transient acceleration of our universe?}",
    eprint = "2008.09602",
    archivePrefix = "arXiv",
    primaryClass = "gr-qc",
    doi = "10.1142/S0218271822500365",
    journal = "Int. J. Mod. Phys. D",
    volume = "31",
    number = "05",
    pages = "2250036",
    year = "2022"
}

@article{Escobal:2023,
    author = "Escobal, A. A. and Jesus, J. F. and Pereira, S. H. and Lima, J. A. S.",
    title = "{Can the Universe decelerate in the future?}",
    eprint = "2302.01946",
    archivePrefix = "arXiv",
    primaryClass = "astro-ph.CO",
    doi = "10.1103/PhysRevD.109.023514",
    journal = "Phys. Rev. D",
    volume = "109",
    number = "2",
    pages = "023514",
    year = "2024"
}

@article{Fortunato:2024,
    author = "Fortunato, J. A. S. and Hipolito-Ricaldi, W. S. and Videla, N. and Villanueva, J. R.",
    title = "{Cosmic slowing down of acceleration with the Chaplygin{\textendash}Jacobi gas as a dark fluid?}",
    eprint = "2406.13132",
    archivePrefix = "arXiv",
    primaryClass = "gr-qc",
    doi = "10.1140/epjc/s10052-025-13996-3",
    journal = "Eur. Phys. J. C",
    volume = "85",
    number = "3",
    pages = "274",
    year = "2025"
}

@book{dodelson2021modern,
  title={Modern cosmology},
  author={Schmidt, F. and Dodelson, S.},
  year={2021},
  publisher={London: Academic Press.}
}

@book{piattella2018lecture,
  title={Lecture notes in cosmology},
  author={Piattella, Oliver},
  year={2018},
  publisher={Springer},
  DOI = "https://doi.org/10.1007/978-3-319-95570-4"
}

@book{carrollspacetime,
    author = "Carroll, Sean M.",
    title = "{Spacetime and Geometry}",
    isbn = "978-0-8053-8732-2, 978-1-108-48839-6, 978-1-108-77555-7",
    publisher = "Cambridge University Press",
    month = "7",
    year = "2019"
}

@misc{NIST:DLMF,
  author       = {{NIST Digital Library of Mathematical Functions}},
  title        = {NIST Digital Library of Mathematical Functions},
  year         = {2023},
  url          = {https://dlmf.nist.gov/},
  note         = {Release 1.1.10}
}

@article{Guedezounme:2025wav,
    author = "Guedezounme, S{\^e}cloka L. and Dinda, Bikash R. and Maartens, Roy",
    title = "{Phantom crossing or dark interaction?}",
    eprint = "2507.18274",
    archivePrefix = "arXiv",
    primaryClass = "astro-ph.CO",
    doi = "10.1088/1475-7516/2026/01/062",
    journal = "JCAP",
    volume = "01",
    pages = "062",
    year = "2026"
}

@article{DESI:2024aqx,
    author = "Calderon, R. and others",
    collaboration = "DESI",
    title = "{DESI 2024: reconstructing dark energy using crossing statistics with DESI DR1 BAO data}",
    eprint = "2405.04216",
    archivePrefix = "arXiv",
    primaryClass = "astro-ph.CO",
    doi = "10.1088/1475-7516/2024/10/048",
    journal = "JCAP",
    volume = "10",
    pages = "048",
    year = "2024"
}

@article{Malekjani:2024bgi,
    author = "Malekjani, Mohammad and Davari, Zahra and Pourojaghi, Saeed",
    collaboration = "DESI",
    title = "{Cosmological constraints on dark energy parametrizations after DESI 2024: Persistent deviation from standard {\ensuremath{\Lambda}}CDM cosmology}",
    eprint = "2407.09767",
    archivePrefix = "arXiv",
    primaryClass = "astro-ph.CO",
    doi = "10.1103/PhysRevD.111.083547",
    journal = "Phys. Rev. D",
    volume = "111",
    number = "8",
    pages = "083547",
    year = "2025"
}

\end{document}